\documentclass[aps,prd,letterpaper]{revtex4}

\usepackage{latexsym}
\usepackage{epsfig}
\usepackage{scrextend}
 \usepackage[french]{babel}

\usepackage{url}
\usepackage{hyperref}
\usepackage{latexsym}
\usepackage{epsfig}
\usepackage{scrextend}
\usepackage{lipsum}   
\usepackage{graphicx}
\usepackage{pdfpages}
\usepackage[title]{appendix}

\usepackage{natbib}

\usepackage{tikz}
\usetikzlibrary{positioning,arrows,decorations.pathmorphing,arrows.meta}

\usepackage{xcolor}
\usepackage{amssymb}
\usepackage{amsmath}
\usepackage{graphicx}

 \definecolor{blue}{RGB}{7,80,201}
 \definecolor{red}{RGB}{200,20,1}

\def\be{\begin{equation}}
\def\ee{\end{equation}}
\def\bea{\begin{eqnarray}}
\def\eea{\end{eqnarray}}
\def\ba{\begin{array}} 
\def\ea{\end{array}}
\def\bc{\begin{center}}
\def\ec{\end{center}}

\def\ghost#1{}

\def\simge{\mathrel{%
   \rlap{\raise 0.511ex \hbox{$>$}}{\lower 0.511ex \hbox{$\sim$}}}}
\def\simle{\mathrel{
   \rlap{\raise 0.511ex \hbox{$<$}}{\lower 0.511ex \hbox{$\sim$}}}}

\def\dis{\displaystyle}

\begin{document}
 
\vspace*{5mm}
\title{\boldmath \large  

\framebox [12.2cm]{\rule[-.85cm]{0cm}{1.95cm} $ \dis
\ba{c}
\hbox{The Yukawa potential of a non-homogeneous sphere,}
\vspace{3mm} \\
\hbox{with new limits on an ultralight boson} 
\ea  $}
\vspace{5mm}\\
}

\author{{\sc P}{ierre} {\sc Fayet}
\vspace{4mm} \\ \small }

\affiliation{Laboratoire de physique de l'\'Ecole normale sup\'erieure\vspace{.5mm}\\ 
ENS-PSL, CNRS, Sorbonne Univ., Univ.\,\,Paris Cit\'e, Paris, France
\footnote{pierre.fayet@phys.ens.fr}
\vspace{.5mm}\\
\hbox{and \,Centre de physique th\'eorique, \'Ecole polytechnique, IPP, Palaiseau, France}
\vspace{5mm}}

\begin{abstract}

\textwidth 10cm

{\vspace{10mm}
\hspace*{2mm}Extremely weak long-range  forces 
\vspace{.2mm} 
may lead to apparent violations of the Equivalence Principle.
The final {\it MICROSCOPE\,}  result, leading at 95\,\% CL to $|\delta| \!< 4.5 \times 10^{-15}$ or $6.5 \times 10^{-15}$ for a posi\-tive or negative E\"otv\"os parameter $\delta$,  requires taking into account the spin of the mediator,  and the sign of $\Delta (Q/A_r)_{\rm Ti\hbox{\scriptsize -}Pt}$ ($Q$ denoting the new charge involved).
\vspace{0mm}
A coupling to $B\!-\!L$ or $B$ should verify   $ |g_{B-L}| < 1.1 \times 10^{-25}$ or 
$|g_{B}| <  8 \times 10^{-25}$, for a spin-1 mediator of mass $m < 10^{-14}$ eV$/c^2$, 
with slightly different limits of $1.3 \times 10^{-25}$ or $\,6.6 \times 10^{-25}$ in the spin-0 case.
\vspace{1.3mm}\\
\hspace*{2mm} 
The limits increase with $m$, in a way which depends on the density distribution within the Earth. This  involves an hyperbolic form factor, 
expressed through a  bilateral  Laplace transform as 
\hbox{$\Phi(x=mR)$}  $= \langle\,\sinh mr/mr\, \rangle$,  
\vspace{-.2mm}
related by analytic continuation to the Earth form factor $\Phi(ix)=  \langle \,\sin mr/mr \,\rangle $.  \,It may be expressed as 
 $\,\Phi(x) = \frac{3}{x^2}\,(\cosh x - \frac{\sinh x}{x}) \times \bar\rho(x)/\rho_0$, where $\bar\rho(x)$ is an effective density, 
\vspace{-.1mm}
decreasing from the average $\rho_0$ at $m=0$ down to the density at the periphery. 
\vspace{1.3mm}\\
\hspace*{2mm} 
\,We give general integral or multishell expressions of $\Phi(x)$, 
\vspace{-.8mm}
evaluating it, and $\bar\rho(x)$, in a simplified 5-shell model.
\vspace{-.8mm}
$\Phi(x)$ may be expanded as $\, \sum \frac{x^{2n}}{(2n+1)!} \frac{\langle \,r^{2n}\,\rangle}{R^{2n}}\!
 \simeq 1 \,+\, .0827\ x^2 +\, .00271\, x^4 +\, 4.78 \times \!10^{-5}\,x^6 + \,5.26\times\! 10^{-7} \, x^8
+\,  ...\, $, absolutely convergent for all $x$ and potentially useful up to $x\approx 5\,$. \,The coupling limits
increase at large $x$ like $mR \ e^{mz/2}/\sqrt{1+mr}\,$
($z=r-\!R$ being the satellite altitude), getting multiplied by $\simeq 1.9,\, 34$, or $1.2\times 10^9$,  
for $m = 10^{-13},\,10^{-12}$ or $10^{-11}$ eV$/c^2$, respectively.
}
\end{abstract}

\maketitle
 
\newpage
\large

\section{Introduction}

The identity between the inertial and gravitational masses of a particle, well tested in the past by the E\"otv\"os experiment
\cite{eot}, is one of the most fundamental principles of physics, and a building block of general relativity. 
The expected limitations of the latter, especially when trying to include gravitational interactions into a quantum framework, 
as well as the search for new feeble interactions weaker than gravity, have motivated further tests of this Equivalence principle \cite{adel,adelbis,ew,microold,micronew}.
These experiments provide strong constraints on very weak new long range forces, usually taken as acting effectively, with strength 
$\epsilon \,e$,  on a charge $Q$ such as $B,\,B\!-\!L$ or $L$ \cite{U}.
They could  lead to apparent violations of the equivalence principle through the elementary interaction potential
\be
\label{V}
V_{ij}(r_{ij}) = \pm \ \epsilon^2 \alpha\ Q_i\,Q_j\, \frac{e^{-r_{ij}/\lambda}}{r_{ij}}\,=\,\pm\ \epsilon^2\, \frac{\alpha}{G_N u^2}\ \left(\frac{Q}{A_r}\right)_{\!i}\,\left(\frac{Q}{A_r}\right)_{\!j}\, G_N\,m_i\,m_j\,\frac{e^{-r_{ij}/\lambda}}{r_{ij}}\ .
\ee
$Q$ and $A_r$ denote the new charge and relative atomic mass (in terms of the atomic mass unit $u$) of the two interacting bodies considered,
the $+$ and $-$ signs are associated with spin-1 and spin-0 mediators, respectively, and the prefactor $\,\pm\ \epsilon^2\, \alpha/G_N u^2
\simeq \pm\, 1.25 \times 10^{36}\, \epsilon^2$, often denoted as $\alpha_5$, has to be much smaller than 1 in magnitude if the new force, when supposed to be long-ranged,  is to be much smaller than gravity.

\vspace{3mm}

The first result of the {\it MICROSCOPE} experiment \cite{microold} led to the constraints \cite{fayetmicro}
\be
\label{limold}
|\epsilon_B|\, <  \,5 \times 10^{-24},\ \ \,|\epsilon_{B-L}| \, <  \,8.4 \times 10^{-25},\  \ \,  |\epsilon_{L}|  \, < \,8.6 \times  10^{-25},
\ee
at the $2\sigma$ level, for a long range force coupled to $B,\,B-L$, or $L$, with strength $\epsilon\,e$.
Its  final result, with a gain in sensitivity by a factor 4.6, leads to an  improvement of these limits by about $2$,
roughly down to 
\be
\label{limnew}
|\epsilon_B|\, <  \,2.5 \times 10^{-24}\,,\ \ \ |\epsilon_{B-L}| \ \hbox{or} \   |\epsilon_{L}|  \, < \, 4 \times  10^{-25}\,,
\ee
or, for couplings $g=\epsilon \,e$ to $B$, $B-L$, or $L$,
\be
\label{limnew2}
|g_B| \, < \,8\times 10^{-25}, \ \ \  |g_{B-L}| \ \hbox{or} \   |g_{L}|  \,< \, 1.2 \times  10^{-25}\,.
\ee

\vspace{2mm}

To better estimate these limits, we need to take into account the dissymmetry of the 
final {\it MICROSCOPE\,} result on the E\"otv\"os parameter, 
$\,\delta_{\hbox{\small \,Ti-Pt}}=(-1.5\,\pm\, 2.3_{\text{\,stat}} \,\pm \,1.5_{\text{\,syst}})\,\times 10^{-15}$
\cite{micronew}.
This implies slightly more restrictive limits on  the magnitude of $\delta$, and thus on the $|\epsilon|$'s, if  $\delta$ is supposed to be positive, 
rather than negative.
The expected sign of $\delta_{\rm\,Ti\hbox{-}Pt}$ depends on the spin of the new force mediator,
and on the sign of the difference $\Delta (Q/A_r)_{\rm Ti\hbox{-}Pt}$.

\vspace{2mm}
Extending  the standard model gauge group to an extra $U(1)$, generated by a combination of the weak hypercharge  $Y$ 
with baryon and lepton numbers $B$ and $L_i$, leads to a new spin-1 boson $U$, possibly very light and very weakly coupled \cite{U}. 
It appears as a generalized dark photon, coupled to a combination of electromagnetic with $B$ and $L_i$ currents.  \linebreak
Axial couplings, that may lead to new forces acting on particle spins  \cite{fa86} and possibly $CP$-violating monopole-dipole interactions \cite{CQG}, may also be present but play no role here. The vector coupling of the new $U$ boson to $Q_{\rm el}$, $B$ and $L_i$ may be parametrized as
\be
\label{eps}
(\epsilon_{Q_{\rm el}} \,Q_{\rm el} + \epsilon_B B + \epsilon_{L_i} L_i)\,,  \ \hbox{times} \  e=\sqrt{4\pi\alpha}\simeq .3028\ .
\ee
Acting on neutral matter proportionally to $\epsilon_B B + \epsilon_{L_i} L_i$, or $\epsilon_{B-L}\,(B-L)$ in a grand-unified theory, the new force
(of range $\lambda \simeq $ $197.327\, \hbox{km} \times [m/(10^{-12} \ \hbox{eV}/c^2)]^{-1}\,$ where $m$ is the new boson mass) may lead to very small apparent violations of the Equivalence Principle.
Such an ultralight boson is now frequently considered, also, as a possible dark matter candidate, and many experiments 
are trying to detect its effects \cite{ligo1,adel2,ppta,nano,yu,fre,all,sun,ama,ligo2,oku,an,ama2}. 
Ultralight spin-0 bosons may also be considered   \cite{dp,kw,dd}, although there is no reason to expect a coupling 
to a combination of $Q_{\rm el}$ with $B$ and $L_i$ in this case.

\vspace{2mm}

We shall discuss couplings to $B,\,L$, or $B\!-\!L$ (or $L_e-L_\mu$ or $L_e-L_\tau$,  or to a combination of them with the electric charge $Q_{\rm el}$), for both spin-1 and spin-0 mediators.
As
\be
\label{Delta00}
\Delta (B/A_r)_{\rm Ti-Pt} \simeq \,0.00079\,,\ \ \Delta (L/A_r)_{\rm Ti-Pt} \simeq \,0.05704\,, \ \  \Delta ((B-L)/A_r)_{\rm Ti-Pt} \simeq\,-\,0.05625\,,
\ee
$\delta_{\rm\,Ti\hbox{-}Pt}$ is expected negative for a spin-1 induced force coupled to $B$ or $L$, and positive if coupled to $B-L$; 
and conversely for a spin-0 induced force.
A positive $\delta$ being slightly more constrained than a negative one in view of the dissymmetry of the final result, 
{\it a force coupled to $B\!-\!L$ gets slightly more strongly constrained  if induced by a spin-1 particle, rather than by a spin-0 one.}
The situation is opposite for couplings to $B$, or $L$.

\vspace{2mm}

The E\"otv\"os parameter $\delta$, expressed in terms of the $\epsilon$ parameters in (\ref{V},\ref{eps}), is given 
in the ultralight or massless case by \cite{fayetmicro}
\vspace{-6mm}

\be
\label{deltaeps0}
\delta_{\hbox{\footnotesize Ti-Pt}}\,= \ \mp\! \!\underbrace{\ \ \frac{\alpha}{G_N\,u^2}\ \ }_{\hbox{\normalsize$1.2536\times 10^{36}$}}\! \!\epsilon^2 \, \left(\hbox{\normalsize $\dis \frac{Q}{A_r}$}\right)_\oplus 
\Delta \! \left(\hbox{\normalsize $\dis \frac{Q}{A_r}$}\right)_{\rm Ti-Pt}
\simeq \ 
\left\{\ \ba{lcc}
\mp\  \  1.00       & \!  \times\  10^{33}  \!  &   \epsilon_B^2\,,
\vspace{1mm}\\
\pm\      3.623    &  \!  \times \ 10^{34}  \!  &   \epsilon_{B-L}^2\,,
\vspace{1mm}\\
\mp\       3.482  & \!   \times \  10^{34} \!   &  \epsilon_L^2\,.
\ea\right.
\ee
$\epsilon_L$  is relative to the electronic number $L_e$, in case of a non-universal coupling to lepton numbers, as for $L_e-L_\mu$ or $L_e-L_\tau\,$.
\,The upper signs correspond to a spin-1 $U$ boson (or generalized dark photon)  mediator, and the lower ones to a spin-0 mediator. 
The first {\it MICROSCOPE\,} result \cite{microold},
$\delta_{\hbox{\footnotesize Ti-Pt}}=\,(-.1\pm .9_{\text{\,stat}} \pm .9_{\text{\,syst}})\times 10^{-14}$,
\vspace{.5mm}
 implies at the $2\sigma$ level
\be
\label{micro2}
|\delta_{\hbox{\footnotesize Ti-Pt}}|_{\rm first} < \,2.55\times 10^{-14}\ \ \ \ (2\,\sigma)\,,
\ee
providing from (\ref{deltaeps0}) the limits (\ref{limold}) on the $\epsilon$'s.
With $|\Delta (B/A_r)_{\rm Ti-Pt}|$ about 72 times smaller than $|\Delta (L/A_r)|$ and $|\Delta( (B-L)/A_r)|$,
and roughly similar numbers of protons, electrons and neutrons within the Earth, 
the limits are similar for couplings to $L$ or $B-L$, and about 6 times stronger than for a coupling to $B$, as shown by (\ref{limold},\ref{limnew},\ref{deltaeps0}).

\section{\boldmath A dissymmetric experimental result}
\label{sec:resc}

The final {\it MICROSCOPE\,} result has improved the precision down to 
$\delta_{\hbox{\small \,Ti-Pt \,\text {final}}}=\,(\,-1.5\pm 2.3_{\text{\,stat}} \pm 1.5_{\text{\,syst}}\,)\times 10^{-15}$  \cite{micronew}.
The corresponding uncertainty, still conservatively taken at the $2\sigma$ level, decreases from the initial $25.5 \times 10^{-15}$ in (\ref{micro2}) down to 
\be
2\sigma
\,\simeq \, 5.5\times 10^{-15}\,,
\ee
with a gain in precision by about 4.6\,. The global constraint on $\delta$ gets reinforced from (\ref{micro2})
down to 
$ -\,7 \times 10^{-15} < \delta_{\hbox{\small \,Ti-Pt}}<4\times 10^{-15}$ at $2\sigma$.
Or, assuming a Gaussian distribution of the result, at a global 95\,$\%$ confidence level 
corresponding to about $1.96 \  \sigma  \simeq 5.4\times 10^{-15}$,
\be
\label{micronew2}
-\ 6.9 \times 10^{-15} <\,\delta_{\hbox{\small \,Ti-Pt}} <\, 3.9\times 10^{-15}\ \ \ (\hbox{global  95\,\%  CL})\,.
\ee
In the absence of the central value $\delta_\circ = -\,1.5 \times 10^{-15}$
one would have $|\delta |< 5.4 \times 10^{-15}$ at the $95\,\%$ CL. 
The negative $\delta_\circ$, however,  leads for $\delta >0$ to a {\it decrease} of the 95\,\% CL limit on $|\delta|$ 
below $5.4\times 10^{-15}$, and  for  $\delta <  0$ to a corresponding {\it increase} above $5.4\times 10^{-15}$.

\vspace{2mm}

Furthermore the asymmetry of the interval (\ref{micronew2}) allowed at the global $95\,\%$ level, when the sign of $\delta$ is kept free, leads for $\delta >0$ to a 95\,\% CL  limit {\it somewhat larger than the too strict} $\,3.9 \times 10^{-15}$ in (\ref{micronew2});
and conversely, for $\delta<0$, to an upper limit on $|\delta|$ {\it somewhat smaller than $\,6.9 \times 10^{-15}$}.
The upper limits on $|\delta|$ 
\vspace{1.5mm}
should then satisfy, at the $95\,\%$ CL,
\be
\label{limlimdelta}
\left\{\ 
\ba{rccr}
\hbox{for} \ \delta > 0\,,&\ \ 3.9 \times 10^{-15}<&\ \hbox{lim}\, \delta \ & < \,5.4 \times 10^{-15} \,,
\vspace{3mm}\cr
\hbox{for} \ \delta < 0 \,,&\ \ 5.4 \times 10^{-15}< &\hbox{lim}\, |\delta| &  <\,6.9  \times 10^{-15} \,.
\ea
\right.
\ee
We thus obtain for  $\delta>0$ a gain in precision, compared with the first result, between $25/5.4 \simeq 4.6\,$ and $25/3.9 \simeq 6.4\,$, 
leading to an improvement of the limits  by {2.15 to 2.5} ; and for $\delta< 0$ to a gain in precision 
between $25/6.9 \simeq 3.6\,$ and $25/5.4 \simeq 4.6$, for an improvement of the limits by {1.9 to 2.15}.
We then get from (\ref{deltaeps0},\ref{limlimdelta}), for a sufficiently light mediator of mass $< \ 10^{-14}$ eV/$c^2$,  
\be
\label{limnew3}
\left\{\, 
\ba{cccr}
\hbox{if} \ \delta >0 :\  &|\epsilon_B| < \, [\ 2 \ \, \hbox{to}\ \,2.3\ ] \times 10^{-24}\, ,&\  |\epsilon_{B-L}|  \ \hbox{or}\  |\epsilon_{L}|  <  \,[\,3.3 \ \,\hbox{to}\,\ 3.9\,]
 \times  10^{-25}\,,
\vspace{2mm}\\
\hbox{if} \ \delta <0 : \ &|\epsilon_B|< \, [\,2.3  \ \hbox{to}\ 2.6\,] \times 10^{-24},&\  |\epsilon_{B-L}|  \ \hbox{or}\  |\epsilon_{L}| <  \,[\,3.9 \ \,\hbox{to}\,\ 4.4\,] \times  10^{-25}\,,
\ea \right.
\ee
at a $95\,\%$ CL.
In particular for a spin-1 induced force coupled to $B\!-\!L$ or $B$, one has
\be
\label{limnewbl}
|g_B| = \,e\, |\epsilon_B|\,<\, [\,7\ \,\hbox{to}\ \,8\,] \times  10^{-25}\,,\ \ \   |g_{B-L}| = \,e\, |\epsilon_{B-L}|\,<\, [\,1\ \,\hbox{to}\ \,1.2\,] \times  10^{-25}\,.
\ee
These naive estimates will be confirmed by a more detailed analysis, leading to eqs.\,(\ref{lfinal41},\ref{lfinal40}).

\section{Precise determination of the limits}
\label{sec:dis}

\subsection{\boldmath  {\em MICROSCOPE} \,limits on the E\"otv\"os parameter $\delta$}

\vspace{-2mm}

\begin{figure}
\caption{\ 
 The {\it MICROSCOPE\,} result 
 \vspace{-.7mm}
$\,\delta_{\hbox{\scriptsize \,Ti-Pt}}=\,(\,-1.5\pm 2.3_{\text{\,stat}} \pm 1.5_{\text{\,syst}}\,)\times 10^{-15}$,  represented proportionally to $\,e^{-(\delta - \delta_0)^2\!/\,2\sigma^2}$  (in blue), \,with $\delta_0 \simeq  -1.5\times 10^{-15}$ and $\sigma\simeq 2.75 \times 10^{-15}$. It leads for an unconstrained $\delta$ to $\,-\,7\times 10^{-15}\!< \,\delta< 4\times 10^{-15}$ at the $2\sigma$ CL (versus $\,-\,5.5\times 10^{-15}\!< \,\delta< 5.5\times 10^{-15}$ for $\delta_0=0$, cf.~dashed curve in orange).
The negative $\delta_0$ leads to limits on $|\delta|$ {\it increased} to about $4.5 \times 10^{-15}$
for $\delta>0$, and {\it decreased} to $6.5 \times 10^{-15}$ for $\delta < 0$, at the 95\,\% CL. 
\ 29.24\,\% of the area under the blue curve corresponds to $\delta >0$  (including 
1.462\,\% for $\delta > 4.5\times 10^{-15}$), and 70.76\,\% for $\delta < 0$ (including 3.538\,\% for $\delta < -\,6.5 \times 10^{-15}$).
\vspace{5mm}}
\label{gauss4}
\includegraphics[width=9.5cm,height=6.2cm]{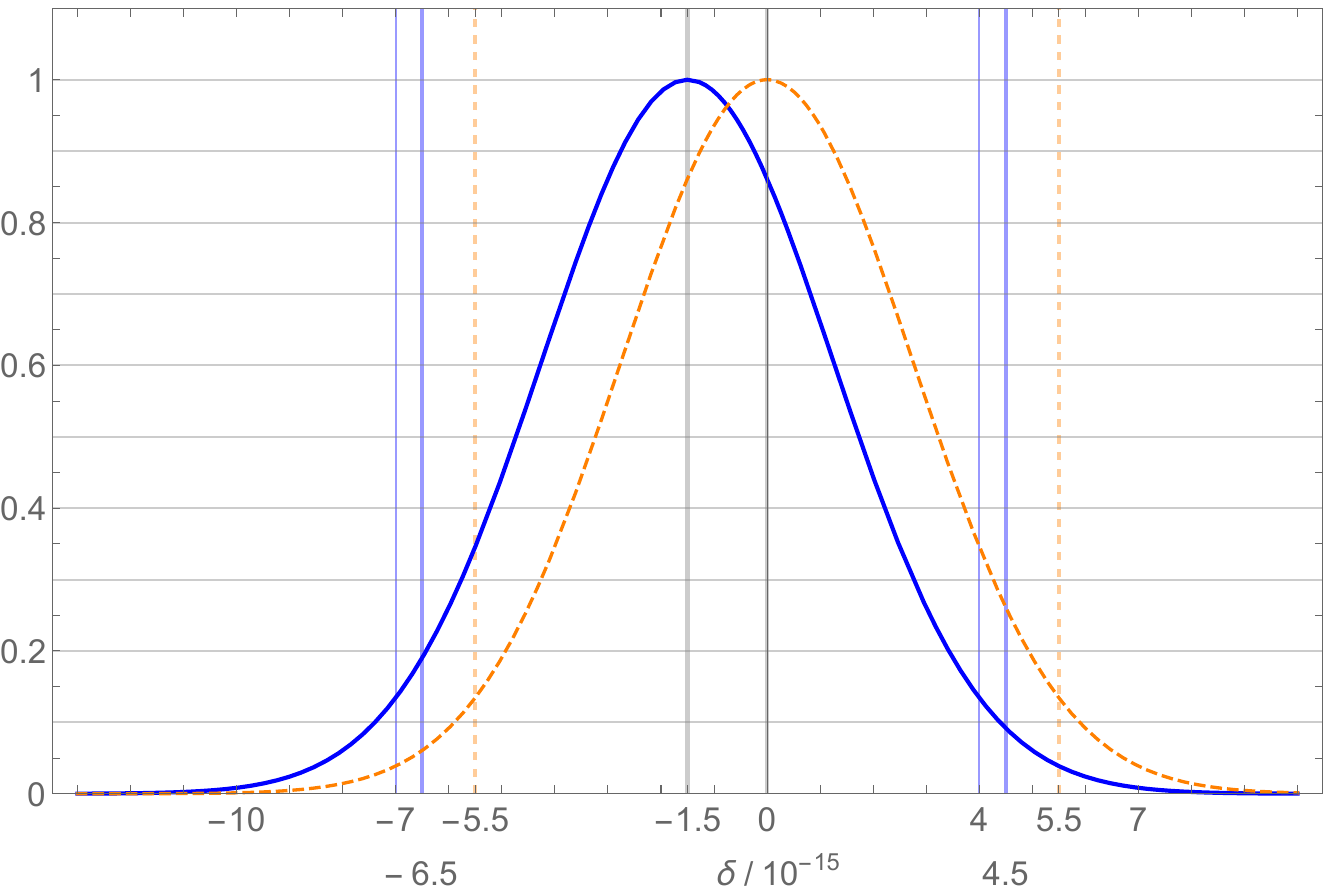}
\vspace{2mm}
\end{figure}

The allowed range of values for $\delta$ is given,  at the global $95\,\%$ confidence level, by the dissymmetric interval  (\ref{micronew2}).
However, the new forces that we are looking for lead, in each case considered, to a $\delta$ of a specific sign.
The best way of taking this into account when estimating confidence levels may be debatable, and we need additional assumptions, e.g. by considering the possible distribution associated with the experimental result to be Gaussian, as represented in Fig.\,\ref{gauss4}. This 
hypothesis has, fortunately, little effects on the results, as seen from eqs.\,(\ref{limlimdelta}-\ref{limnewbl}). 

\vspace{2mm}
Let us consider a Gaussian probability distribution expressed as 
\be
\label{p}
p(\delta/\sigma) = \frac{e^{-(\delta - \delta_0)^2\!/\,2\sigma^2}}{\sqrt{2\pi}}\,.
\ee
\vspace{-3mm}

\noindent We take
erf$({x}/{\sqrt 2})\!=\!\int_{-x}^x \, \frac{1}{\sqrt{2\pi}}\ e^{-\frac{t^2}{2}}\, dt$
\vspace{-.5mm}
as the probability for $\,t=(\delta-\delta_\circ)/\sigma$ 
to lie between $-x\sigma$ and $x\sigma$, 
\,95\,\% or 90\,\% confidence levels corresponding to $x\sigma\simeq 1.960 \ \sigma$  or $1.645\ \sigma$.
With $\delta_\circ = -\,1.5\times 10^{-15}$, $\sigma\simeq 2.7459\times 10^{-15}$,
$|\delta_\circ|/(\sigma \sqrt 2) \simeq$ $ .38627$, and 
\hbox{\,erf}\,$(|\delta_\circ|/(\sigma \sqrt 2) \simeq .41512\,$,
about 41.51 \% of the probability distribution corresponds to $\delta$ between $-\,2\, |\delta_\circ| $ and $0$. This implies  a 29.24 \% probability  for $\delta >0$\,,
i.e.,
\be
\frac{1-\hbox{erf }(|\delta_\circ|/(\sigma \sqrt 2))}{2}\ = \ \int_{|\delta_\circ|/\sigma}^\infty \,\frac{1}{\sqrt{2\pi}}\ e^{-\frac{t^2}{2}}\, dt\,\simeq \,29.24\ \%\ ,
\ee
and the remaining 70.76 \% for $\delta <0$\,.

\vspace{2mm}
Integrating $p(\delta/\sigma)$ in eq.\,(\ref{p}) we find the upper limits on $|\delta|$ by arranging, for $\delta >0$,  
for having 5\,\% of the 29.24\,\% area under the blue curve in Fig.\,\ref{gauss4} in the positive $\delta$ region (i.e.,~1.462\,\% of the total area) 
above the upper limit of $\simeq 4.5 \times 10^{-15}$. 
And similarly,  for $\delta< 0$, for having 5\,\% of the 70.76\,\% area in the negative $\delta$ region (i.e.,~3.538\,\% of the total) 
under the limit of $-\,6.5\times 10^{-15}$.
Altogether we get
\vspace{-4mm}

\be
\label{limdelta}
\framebox [10.5cm]{\rule[-.65cm]{0cm}{1.55cm} $ \dis
\left\{\ 
\ba{crlc}
 \hbox{for $\delta >0$ ,} \ & \delta \,< \,4.5 \times 10^{-15}\, &  
\vspace{2mm}\cr
 \hbox{for $\delta <0$ ,}  \ & |\delta| < \,6.5  \times 10^{-15}\,&
\ea
\right.\ \ \ \hbox{at the 95\,\% CL},
$}
\ee
as illustrated in Fig.\,\ref{gauss4}, in place of the $[ -\,6.9\times 10^{-15},\, 3.9 \times 10^{-15}]$  interval
in (\ref{micronew2}) and in agreement with (\ref{limlimdelta}).
These limits come in place of the earlier $|\delta| < 25 \times 10^{-15}$ at the 95\,\% CL from the first {\it MICROSCOPE\,} result.

\subsection{Influence of the confidence level on the limits}

If $\delta$ is positive it has about 2.5\,\%\,/29.24\,\% $\simeq$ 8.55 \% chances to be above the 
$3.9\times 10^{-15}$ in (\ref{micronew2}), which provides an upper limit at the $\simeq 91.45\,\%$ CL.
To get an upper limit, expressed as $\delta_\circ\! + x\sigma$, on a positive $\delta$ at a given confidence level $\,cl$, 
we express that the probability for a positive $\delta$ to be larger than $\delta_0 +x\sigma$ is $1-cl\,$ 
(e.g., 5\,\%, for a 95\,\% CL limit).
This one  is taken as the probability 
for an unconstrained $\delta$ to be larger than $\delta_0 + x\sigma$, 
divided by the probability for $\delta$ to be found positive, here 29.24\,\%.
This reads
\be
\label {P}
{\cal P} \,(\,\delta_{\rm {positive}} > \delta_0+x\sigma\,) \ = \frac{ \int_{x}^\infty \,\frac{1}{\sqrt{2\pi}}\ e^{-\frac{t^2}{2}}\, dt}{\int_{-\,\delta_0/\sigma}^{\ \infty} \,\frac{1}{\sqrt{2\pi}}\ e^{-\frac{t^2}{2}}\, dt}\ =\ 1-\ cl\ 
\ee
(this ratio ${\cal P}$ is of course 1, for $\delta_0 + x\sigma = 0$), so that
\be
\label{erf1}
\int_x^{\infty} \, \frac{1}{\sqrt{2\pi}}\ e^{-\frac{t^2}{2}}\, dt \,=\,\frac{1-\,\hbox{erf}\,(x/\sqrt2)}{2} \,= \, 
\underbrace{\int_{-\,\delta_0/\sigma}^{\, \infty} \, \frac{1}{\sqrt{2\pi}}\ e^{-\frac{t^2}{2}}\, dt}_{\hbox{\small $\simeq$ 29.24 \%}}\  \times \ (1-cl) \,,
\ee
\vspace{-6mm}

\noindent
which leads to\,\,  \footnote{For a symmetric distribution the above 29.24\,\% would be replaced by the 50\,\% probability for an unconstrained $\delta$ to be positive, eq.\,(\ref{erf2}) simplifying into $ x=\sqrt 2\ \, \hbox{erf}^{-1}(cl)$, providing the usual  90\,\% and 95\,\% CL  limits at
$x\sigma \simeq 1.645 \ \sigma$ and 1.960 $\sigma$. 
}

\vspace{-6mm}

\be
\label{erf2}
x \,\simeq\,\sqrt 2\ \,\hbox{erf}^{-1}[ \,1- 2\times 29.24\,\%\ (1-cl)\,]\,.
\ee

\noindent
The resulting limits at the $95\,\%, 91.45\,\%$ and $90\,\%$ confidence levels are  shown in Table~\ref{table:lim+}.

\begin{table}
\caption{\ Upper limits on the E\"otv\"os parameter $\delta$, when positive. Expressed as 
$\delta_\circ+x\sigma$ with $\delta_\circ = -1.5\times 10^{-15}$ and $ \sigma \simeq 2.75\times 10^{-15}$,  they are given at the  $95\,\%, \,91.45\,\%$ and $90\,\%$ confidence levels.
The resulting 90\,\% CL limits on the $|\epsilon|$'s are smaller than at the 95\,\% level by a factor 
$\simeq \sqrt{4.5/3.7}\simeq 1.10$.}
\label{table:lim+}
\vspace{2mm}
\normalsize
$\ba{|c||c|c|c|}
\hline  &&& \\ [-4mm]
 \hbox{conf. level}& 95\,\% & 91.45\,\%& 90\,\%
\\ [1mm]
\hline
&&& \\  [-2mm]
x &\ \sqrt 2 \ \hbox{erf}^{-1}(0.97076)\ &\ \  \sqrt 2\  \hbox{erf}^{-1}(0.95) \ \ & \ \sqrt 2\ \hbox{erf}^{-1}(0.94152)\ 
\\ [0.8mm]
 &\simeq \,2.18& \simeq \,1.96 & \simeq \,1.89
\\ [2mm]
x\sigma &\ \ 6\times 10^{-15} &  5.4 \times 10^{-15} & 5.2 \times 10^{-15}
\\ [1.5mm]
\ \hbox{lim}\ \delta =x\sigma +\delta_\circ \ &4.5 \times 10^{-15} & 3.9 \times 10^{-15}&3.7 \times 10^{-15}
\\ [2mm]
\hline
\ea 
\vspace{-1mm}
$
\end{table}

\begin{table}
\caption{\ Upper limits on $|\delta|$, when $\delta$ negative, expressed as 
$|\delta_\circ|+x\sigma$, and given at the 
96.47~\%, 95\,\% and 90\,\% confidence levels.
\vspace{-.3mm}
The 90\,\% CL limits on the $|\epsilon|$'s are smaller than at the 95\,\% level by a factor 
$\simeq \sqrt{6.5/5.54}\simeq 1.08$.}
\label{table:lim-}
\vspace{3mm}
\normalsize
$
\ba{|c||c|c|c|}
\hline  &&& \\ [-4mm]
 \hbox{conf. level}& 96.47\,\% & 95\,\%& 90\,\%
\\ [1mm]
\hline
&&& \\  [-2mm]
x &\ \sqrt 2 \ \hbox{erf}^{-1}(0.95)\ & \ \sqrt 2\ \hbox{erf}^{-1}(0.92924)\ &\  \sqrt 2\ \hbox{erf}^{-1}(0.85848)\ 
\\ [.8mm]
 & \simeq 1.96&\simeq 1.81 & \!\simeq 1.47
\\ [2mm]
x\sigma &5.4\times 10^{-15} & \ \  5 \times 10^{-15} & 4.04\times 10^{-15}
\\ [1.5mm]
\ \hbox{lim}\  |\delta| =x\sigma + |\delta_\circ|\  &6.9 \times 10^{-15} & 6.5 \times 10^{-15}&5.54 \times 10^{-15}
\\ [2mm]
\hline
\ea
\vspace{6mm}
$
\end{table}

\vspace{2.5mm}

If $\delta$ is negative, it has about 2.5\,\%/70.76\,\% $\simeq$ 3.53\,\% chances to be below the $-\ 6.9\times 10^{-15}$ in (\ref{micronew2},  
$|\delta| <  6.9 \times 10^{-15}$ then providing a limit at the $\simeq 96.47\,\%$ level.
For  a limit  at a confidence level $cl$ we solve for 
\be
\int_{-\infty}^{-x} \, \frac{1}{\sqrt{2\pi}}\ e^{-\frac{t^2}{2}}\, dt \,=\,\frac{1-\,\hbox{erf}\,(x/\sqrt2)}{2} \,\simeq \  70.76 \% \times (1-cl) \,,
\ee
which leads to
$
x \simeq\sqrt 2\ \,\hbox{erf}^{-1}[ \,1- 2\times 70.76\,\%\ (1-cl)\,]\,
$.
The resulting limits at the $96.47\,\%$, $95\,\%$ and $90\,\%$ confidence levels are shown in Table~\ref{table:lim-}.
We get in particular
\be
\left\{\ 
\ba{crlc}
 \hbox{for $\delta >0$ ,} \ & \delta \,< \,3.7 \times 10^{-15}\,&  
\vspace{2mm}\cr
 \hbox{for $\delta <0$ ,}  \ & |\delta| < \,5.6  \times 10^{-15}\,& 
\ea
\right.\ \ \ \hbox{at the 90\,\% CL}.
\ee

\subsection{\boldmath New limits on the couplings, depending on the mediator spin}

The limits on the $\epsilon$'s, found from (\ref{deltaeps0},\ref{limdelta}), are
\be
\label{lfinal}
\left\{\ 
\ba{cclc}
\hbox{if} \ \delta >0 : &|\epsilon_B| <  \,2.1 \times 10^{-24}\ ,\ \ |\epsilon_{B-L}| \ \,\hbox{or} \ \,  |\epsilon_{L}|   < \,3.6 \times 10^{-25}\,,
\vspace{2.5mm}\cr
\hbox{if} \ \delta <0 : &|\epsilon_B| <  \,2.6 \times 10^{-24}\ ,\ \ |\epsilon_{B-L}|   \ \,\hbox{or} \ \,  |\epsilon_{L}|   < \,4.3 \times  10^{-25}\,,
\ea \right. 
\ee
with an improvement by $\simeq (25/4.5)^{1/2}\simeq 2.36$ for $\delta > 0$, and $\simeq (25/6.5)^{1/2}\simeq 1.96$ for $\delta <0$\,.
This leads to the final result
\be
\label{lfinal2}
\!\left\{
\ba{cclc}
\hbox{spin-1 mediator}: &|\epsilon_B| <  \,2.6 \times 10^{-24},\ \, |\epsilon_{B-L}|  <  \,3.6 \times 10^{-25},\ \,  |\epsilon_{L}|   < \,4.3 \times  10^{-25},
\vspace{2.5mm}\cr
\hbox{spin-0 mediator}: &|\epsilon_B| <  \,2.1 \times 10^{-24}, \ \, |\epsilon_{B-L}|  <  \,4.3 \times 10^{-25},\ \,  |\epsilon_{L}|   < \,3.6 \times  10^{-25},
\ea \right.\!\!\!
\ee
at the 95\,\% confidence level  \  \footnote{Strictly speaking for a $\delta$ of a given sign the limits on $|\epsilon_L|$ are larger than for 
$|\epsilon_{B-L}|$ by about 2\,\%, a difference that we shall usually consider as negligible.}.

\vspace{2mm}
For a given sign of $\delta$ the limits on the $\epsilon$'s, and couplings $g=\epsilon e$, are larger for $B$ and $L$ than for 
$B\!-\!L$, by factors $\,\simeq \!\sqrt {36.23}\simeq 6$ and $\sqrt {36.23/34.82}$ $ \simeq 1.02$, respectively, as seen from (\ref{deltaeps0}). 
For a given type of coupling they are larger for $\delta<0$ than for $\delta >0$ by a factor $\simeq\! \sqrt{6.5/4.5}\simeq 1.2$. The 90\,$\%$ CL limits are obtained for $\delta >0$  (resp. $< 0$) from $\delta < 3.7 \times 10^{-15}$ instead of $4.5\times 10^{-15}$ (resp. $|\delta| < 5.6 \times 10^{-15}$ instead of $6.5\times 10^{-15}$),
cf. Tables \ref{table:lim+} and \ref{table:lim-}. They are only very slightly smaller, by factors $\simeq  1.1$ or 1.08
(close to the $(1.960/1.645)^{1/2}\simeq 1.09 $ that would be obtained for a symmetric distribution),
illustrating the modest effect of the chosen confidence level on the coupling limits.

\vspace{2mm}

We finally remember, for a spin-1 $U$ boson somewhat lighter than  $\approx10^{-14}$ eV/$c^2$ effec\-tively coupled to $B,\ B\!-\!L\,$, or $L$,
\be
\label{lfinal41}
\framebox [14.8cm]{\rule[-1.1cm]{0cm}{2.4cm} $ 
\hbox{spin-1 mediator:}\left\{ \ba{ccl}
\ \,|\epsilon_{B}|  \ <  2.6 \times 10^{-24},&\hbox{or}&\ \  |g_{B}|\  <  \, 7.7 \times  10^{-25},
\vspace{2mm}\\
|\epsilon_{B-L}|  <  3.6 \times 10^{-25},& \hbox{or}& |g_{B-L}|  <  \,1.1\times  10^{-25},\ \ (95\,\%\ \hbox{CL})
\vspace{2mm}\\
\ \ |\epsilon_{L}|  \ <  4.3 \times 10^{-25},& \hbox{or}& \ \ \,|g_{L}|  \ <  \,1.3\times  10^{-25},
\ea\right.
$}
\ee
as expected from (\ref{limnew3},\ref{limnewbl}) \footnote{These 95\,\% CL limits of 1.1 and 1.3 $\times 10^{-25}$ in the spin-1 and spin-0 cases for a $B\!-\!L$ coupling, rounded from 1.07 and 1.28  $\times 10^{-25}$, differ by a factor $\simeq \sqrt{6.5/4.5} \simeq 1.20$. And similarly for the 6.4 and 7.7 $\times 10^{-25}$ limits for a coupling to $B$ in the spin-0 and spin-1 cases,  which also differ by the same factor.}. 
This is almost always more constraining than the limit from a torsion-balance search for ultralow-mass  dark matter \cite{adel2} constraining, at its maximum sensitivity near  $m= 8\times 10^{-18}$ eV$/c^2$, $|g_{B-L}| $ to be less than $10^{-25}$ (at 95\,$\%$ CL), assuming that dark matter predominantly consists of ultralow-mass vector bosons coupled to $B-L$\,. For a spin-0 mediator we get
\be
\label{lfinal40}
\hbox{spin-0 mediator:}\left\{ \ba{ccl}
\ \,|\epsilon_{B}| \ <  2.1 \times 10^{-24},&\hbox{or}&\ \  |g_{B}| \ <  6.4 \times  10^{-25},
\vspace{2mm}\\
|\epsilon_{B-L}|  <  4.3 \times 10^{-25},& \hbox{or}& |g_{B-L}|  <  1.3 \times  10^{-25},\ \ (95\,\%\ \hbox{CL})
\vspace{2mm}\\
\ \ |\epsilon_{L}| \  <  3.6 \times 10^{-25},& \hbox{or}& \ \ \,|g_{L}| \  <  1.1 \times  10^{-25} .
\ea\right.
\ee

\vspace{-.5mm}

\noindent
The 90 $\%$ CL limits are about 8\,\% or 9\,\% lower as seen from Tables \ref{table:lim+} and\ \ref{table:lim-}, down to $1 \times 10^{-25}$ 
for a $B\!-\!L$ coupling in the spin-1 case.

\vspace{-1mm}

\section{\boldmath General expression of  the form factor $\,\Phi(x)\,$ as $\dis \ < \frac{\sinh mr}{mr}> $}
\label{secphi}

\vspace{-1mm}

These limits are valid for a mediator mass $m$ smaller than about $10^{-14}$ eV$/c^2$, corresponding to 
a range $\lambda = \hbar /mc \,\simge  20\,000$ km, larger than the diameter of the Earth. 
They still remain approximately valid somewhat above this value, before increasing significantly above $\approx 10^{-13}$ eV$/c^2$, 
corresponding to $ \lambda \approx 2\,000$ km, 
for which only a smaller part of the Earth can act efficiently as a source of the new force. This may be taken into account using a spherical or ellipsoidal-layered Earth model, as done in \cite{adelbis} for $\lambda >$ 1\,000 km.
Furthermore with a satellite orbiting at a mean altitude $z\simeq 710$~km, the strength of a new force of range $\lambda= \hbar /mc$, 
and the resulting E\"otv\"os parameter $\delta$, also include, for the smaller values of $\lambda$,  
an exponentially small factor proportional to $e^{-mz}$, leading to an increase in the coupling limits by a large factor proportional to $e^{mz/2}$.  
\,$\delta$ also includes, for the lower values of $\lambda$ down to $\approx 20$~km, 
\linebreak
a factor $\bar \rho/\rho_0$ decreasing down to $ \simeq 1/2$, taking into account that the effective density $\bar \rho$ of the regions 
at the origin of the new force decreases, near the surface of the Earth, down to about 1/2 of its average density $\rho_0\simeq 5.51$ g/cm$^3$.

\vspace{2mm}

The Earth will still be considered as spherically symmetric 
\vspace{-.4mm}
with a radius $R\simeq 6\,371$ km.
The analysis involves density ratios $\rho(r)/\rho_0$, where $\rho_0 = \int_0^R \rho(r)\,\frac{3r^2dr}{R^3}$ is the average density. 
We shall consider for simplicity that these ratios are approximately the same  for the new charge (denoted by $Q$), and for the mass, 
viewing the new charge distribution as approximately proportional to the mass distribution.

\subsection{Solution of the Poisson-like equation for the potential}

\vspace{-1mm}

 The Yukawa potential of a pointlike new charge $Q$ at the origin, ${\cal V} (r)= \,Q\,e^{-mr}/4\pi r$, \,is replaced, outside the sphere of radius $R$, by
\be
\label{calV}
{\cal V}_{\rm out } (r)=  \,Q\ \frac{e^{-mr}}{4\pi r}\ \Phi(x)\,,
\ee
with  $\,x=mR =R/\lambda$\,. $\,\Phi(x)$ takes into account the extension of the sphere 
and its non-homo\-geneity, as compared to a pointlike source.
The potential within the sphere, ${\cal V}_{\rm in} (r)$,  may be obtained by solving the Poisson-like equation with the new charge density $\rho(r)$,
\be
\label{poiss}
(\,-\,\triangle + m^2\,)\,{\cal V}_{\rm in} (r) \,=\,\dis \rho(r)\ .
\ee
A thin shell of radius $r'$, density $\rho(r')$ and thickness $dr'$ generates, 
at  $r\geq r'$, an outside potential proportional to  $e^{-mr}/r$; and at $r\leq r'$, an inside potential 
proportional to $\sinh mr/mr$. Both are equal at $r=r'$, for which the induced potential is proportional to the product 
$({\sinh mr'}/{mr'})\,({e^{-mr'}}/r')$\,.
\,It is then equal to 
$\rho(r')\, r'^2dr' $ $\,({\sinh mr'}/{mr'})\,({e^{-mr'}}/r')\,$, \,up to a possible proportionality factor found to be 1 as for $m=0$, corresponding to a Coulomb potential $[\,\rho(r') \, 4\pi r'^2dr'\,] /4\pi r'$.

\vspace{2mm}

This shell of radius $r'$ thus generates the elementary potential
\vspace{-3mm}

\be
\label{vthin}
d\,{\cal V } (r) =\, \left\{\, \ba{ccc}
\dis \rho(r')\ \,\frac{\sinh mr'}{mr'}\ \frac{e^{-mr}}{r}\ r'^2 \,dr'\, ,&\ \hbox{for }\ \  r\geq r'\,,
\vspace{2.5mm}\\
\dis \rho(r')\ \,\frac{\sinh mr}{mr}\ \frac{e^{-mr'}}{r'}\ r'^2 \,dr'\, , & \ \hbox{for }\ \ r \leq r'\, .
\ea \right. \hspace{-3mm}
\ee
\vspace{-.5mm}

\noindent
For a sphere of radius $R$ the inside potential ${\cal V}_{\rm in} (r)$ is the sum of the contributions (\ref{vthin}) 
from the inner and outer  parts of the sphere, i.e.,
\be
\label{potge0}
\framebox [14.2cm]{\rule[-.5cm]{0cm}{1.3cm} $ \dis
{\cal V}_{\rm in}(r)\,=\ \frac{e^{-mr}}{r}\int_0^r \rho(r')\ \frac{\sinh mr'}{mr'}\ r'^2\,dr'\, +\,  \frac{\sinh mr}{r} \int_r^R \rho(r')\ \frac{e^{-mr'}}{mr'}\ r'^2\,dr'\ .
$}
\ee
\vspace{0mm}

\noindent
Eq.\,(\ref{poiss}) is indeed satisfied, as
\be
\ba{ccl}
\dis \frac{1}{r}\,\frac{d\ }{dr}\, {r\cal V}_{\rm in}(r) \!&=&\!\dis -\,m\,\frac{e^{-mr}}{r}\!\int_0^r \! \rho(r')\ \frac{\sinh mr'}{mr'}\ r'^2\,dr'\, +\, m\, \frac{\cosh mr}{r}\! \int_r^R \!\rho(r')\ \frac{e^{-mr'}}{mr'}\ r'^2\,dr'\,,
\vspace{2mm}\\
\dis \frac{1}{r}\,\frac{d^2\ }{dr^2}\, {r\cal V_{\rm in}}(r) \!&=&\, \dis m^2\,{\cal V}_{\rm in}(r) -\rho(r)\, e^{-mr}\,( \sinh mr+ \cosh mr)
= m^2\,{\cal V}_{\rm in}(r) -\, \rho(r) \,,
\vspace{-6.5mm}\\
\ea
\ee
\vspace{2mm}

\noindent
so that 
$(\,-\,\triangle + m^2\,)\, {\cal V}_{\rm in}(r)=\rho(r)$.

\subsection{\boldmath Expression of $\Phi(x)$}

In a similar way, the outside potential reads 
\be
{\cal V}_{\rm out}(r) =  \frac{e^{-mr}}{r}\int_0^R \rho(r')\ \frac{\sinh mr'}{mr'}\ r'^2\,dr'\,=
\, \frac{Q}{4\pi }\frac{e^{-mr}}{r} \ \Phi(x)\,,
\ee
\vspace{-3mm}

\noindent
with $\,Q=(4\pi\,R^3/3)\,\rho_0$, \,so that
\be
\label{Phi000}
\framebox [10.8cm]{\rule[-.55cm]{0cm}{1.35cm} $ \dis
\Phi(x)\,= \,\int_0^R\,\frac{\rho(r)}{\rho_0}\  \frac{\sinh mr}{mr}\ 
\frac{3r^2dr }{R^3}
\ =\ \langle\ \frac{\sinh mr}{mr}\ \rangle\,\geq \,1\ .
$}
\ee
\vspace{-.5mm}

\noindent
This can also be expressed, independently of $R$ and $\rho_0$, as
\be
\Phi(x)\,= \,\frac{1}{m}\,\int_0^\infty 4\pi \rho(r) \,\sinh mr\, rdr\ ,
\ee
with $\rho$ normalized to unity through $\int_0^\infty 4\pi \rho(r) \,r^2 dr =1$ (but still considering a sphere of finite radius).
The outside potential  at distance $r$ is indeed the above Yukawa potential, reobtained as
\be
\label{calv1}
{\cal V}_{\rm out}(r)\,= 
 \int \,\rho(r)\ d^3\vec r\,' \,\ \frac{e^{-m |\,\vec r-\vec r\,'\,|}}{4\pi\,|\,\vec r-\vec r\,'\,|}\,=
 \,\langle \ \frac{\sinh mr}{mr}\ \rangle\, \times \  Q\ \frac{e^{-mr}}{4\pi r}\ ,
\ee
as verified using $\,l^2 = (\vec r-\vec r\,')^2= r^2+r'^2-2rr' \cos\theta\,$ so that \,$l\,dl = rr'\,\sin\theta\, d\theta$\,, with $l$ varying from 
$\,r-r'>0\,$ to $r+r'$, and
\be
\label{calv2}
{\cal V}_{\rm out}(r)=
\int_0^R \!\rho(r')\ r'^2dr' \int_0^\pi \frac{1}{2}\  \frac{e^{-ml}}{l}\, \sin\theta \,d\theta 
\,= \int_0^R\! \rho(r')\ r'^2dr'  \ \,\frac{e^{-m(r-r')}-e^{-m(r+r')}}{2\,mrr'}\ .
\ee

\vspace{2mm}

It is interesting to note, in view of future developments, that $\rho(r)$ may be extended to an even function of $r$, so that
\be
\Phi(x\!=\!mR) = g(m)= \, \frac{2\pi}{m} \int_{-\infty}^\infty r\,\rho(r) \ e^{mr}\, dr\ ,
\ee
is equal to $2\pi/m$ times the bilateral Laplace transform of the odd function $r\,\rho(r)$, with $g(0)\! = \Phi(0) =1$.

\vspace{3mm}

$\Phi(x) $ appears, for a finite-range Yukawa interaction, as an isotropic form factor taking into account the extension of the spherical body considered, and its inhomogeneous character. 
$\Phi(x)$ is close to 1 for a long range interaction with $x  \simeq 0$ i.e. $\lambda = 1/m \gg R$, 
for which the internal structure of the sphere plays essentially no role.
For an homogeneous sphere $\Phi(x)$ in eq.\,(\ref{Phi000}) reduces to \vspace{-2mm}\cite{bou,sne,fis0,fis,adel3}
\vspace{-1mm}

\be
\label{phi}
\phi(x) 
\,=\,\frac{3}{x^3}\, \int_0^x \sinh u \ u\,du\,
=\,\frac{3\,(x\,\cosh x - \sinh x)}{x^3}\ ,
\ee

\noindent 
with the potential obtained from (\ref{potge0}) as
\be
\label{vinout}
\left\{\ 
\ba{ccc}
{\cal V} _{\rm in}(r) &=&\,
\displaystyle
\frac{\rho_\circ}{m^2}\ \left[ 1 - (x+1)\ e^{-x}\ \frac{\sinh mr}{mr}  \right] \,,
\vspace{1mm}\\
\displaystyle
{\cal V} _{\rm out} (r)&=&
\displaystyle
\frac{\rho_\circ}{m^3} \ (x\,\cosh x-\sinh x)\  \frac{e^{-mr}}{r}\ .
\ea \right.
\ee
\vspace{-1mm}

\noindent
This may also be found directly, the inside and outside potentials being expressed as $\rho_0/m^2 + A \,\sinh mr/mr$ 
and $B \,e^{-mr}/r$, respectively,  which also leads to eqs.\,(\ref{vinout})\,\,\footnote{
\noindent
The continuity of ${\cal V}$ and ${d{\cal V}}/{dr}$ at $r=R$ is expressed as 
\vspace{-2mm}
$$
{\cal V}_{\rm in}(R)=  
 \frac{\rho_\circ}{m^2 x} \,
[\,x - (x+1) \,e^{-x}\sinh x\,]
=  \frac{\rho_\circ}{2 m^2 x} \, [\,x-1+ (x+1)\,e^{-2x}\,]
= \frac{\rho_\circ}{m^2 x} \, (x \cosh x -\sinh x)\, e^{-x} = {\cal V}_{\rm out}(R)\,,
$$
\vspace{-3.5mm}
$$
-\,\frac{d\,{\cal V}_{\rm in}}{dr}\,(R)= \ \frac{\rho_\circ}{m}  \displaystyle \ \frac{(x+1)\ e^{-x}\ (x\, \cosh x -\sinh x)}{x^2}
\,=\, -\,\frac{d\,{\cal V}_{\rm out}}{dr}\,(R)\,.
$$
}
\footnote{For large $\lambda \gg R$ i.e. small $x= R/\lambda$, $\phi(x)\to 1$ and ${\cal V}_{\rm out}(r) $ tends to be the same Yukawa potential as for a pointlike sphere. As $\,\sinh mr/mr \simeq 1 + x^2r^2/6R^2 $, we also recover 
\vspace{-.5mm}
in this massless limit the same inside potential 
$
\,{\cal V} _{\rm in}(r) =
 \frac{Q}{4\pi R} \,  (\frac{3}{2} -\frac12 \frac{r^2}{R^2})
$
\vspace{.5mm}
as for a uniformly charged sphere.
}.

\section{\boldmath Analytic expansion and continuation of $\Phi(x)$}

$\Phi(x=mR)$ may be viewed as an hyperbolic form factor,
expressed as in (\ref{Phi000}) as the weighted average of $\sinh mr /mr = \sum \, (mr)^{2n}\!/(2n+1)!\,$.  
\,It may be expanded as a power series in $x$, absolutely convergent for all $x$, involving the even moments of the density $\rho(r)$, 
\vspace{-8mm}

\be
\langle \ r^{2n}\,\rangle = \frac{1}{\rho_0\, 4\pi R^3/3}\ \int_0^R \rho(r)\ r^{2n}\,4\pi r^2 dr \ =\int_0^R \,\frac{\rho(r)}{\rho_0}\ \frac{3\,r^{2n+2}\, dr}{R^3}\ ,
\ee 
so that
\be
\label{expphi0-2}
\framebox [13.8cm]{\rule[-1.3cm]{0cm}{2.75cm} $ \dis
\ba{ccl}
\Phi(x) \,=\, \dis \langle \ \frac{\sinh mr}{mr}\ \rangle   &=&\dis \sum_0^\infty\ \frac{x^{2n}}{(2n+1)!}\ \frac{\langle \,r^{2n}\,\rangle}{R^{2n}} \  = \ \sum_0^\infty\ \frac{1}{(2n+1)!}\ \frac{\langle \,r^{2n}\,\rangle}{\lambda^{2n}} 
\vspace{2mm}\\ &=&\ \dis
 1\, + \,x^2\,   \frac{\langle\, r^2\,\rangle}{6 \,R^2} +\, x^4\,   \frac{\langle\, r^4\,\rangle}{120\,R^4} +\,x^6\, \frac{\langle \,r^6 \,\rangle}{5\,040\,R^6} 
+ \, ...\ .
\ea
$}
\ee
\vspace{0.5mm}

\noindent
This may also be reexpressed, with no reference to $R$, in terms of a density distribution (vanishing at large $r$) normalized to 1, as 
\be
\Phi(x)\,=\,\frac{1}{m}\,\int_0^\infty 4\pi\,\rho(r) \,\sinh mr \,r dr\,=\,\sum_0^\infty\ \ \frac{1}{(2n+1)!}\ \frac{\langle \,r^{2n}\,\rangle}{\lambda^{2n}}\ ,
\ee
with
\vspace{-8mm}

\be
\langle \ r^{2n}\,\rangle = \int_0^\infty 4\pi\,\rho(r)\ r^{2n+2}\,dr \ .
\ee

\vspace{2mm}

For a pointlike distribution at the origin $\Phi(x)=1$, 
in agreement with eq.\,(\ref{expphi0-2}), all momenta $\langle \,r^{2n}\,\rangle$ being 0 for $n\geq 1$.
For an homogeneous distribution $\rho=\rho_0$ for $r<R$, for which
\be
\label{momh}
\langle \ r^{2n}\ \rangle
\ =\ \frac{3}{2n+3}\ R^{2n}\ ,
\ee
eq.\,(\ref{expphi0-2}) reads
\be
\label{expphi7}
\phi(x)\,=\, \sum_0^\infty\, \frac{x^{2n}}{(2n+1)!}\ \frac{3}{2n+3} \,=\,1 +\,\frac{x^2}{10}\,+\, \frac{x^4}{280}\, + \, \frac{x^6}{15\,120}\,+ \, \frac{x^8}{1\,330\,560}\, 
+ \ ...\ ,
\ee
which is, not surprisingly, the expansion of
\be
\label{expphi8}
\phi(x) \,=\,\frac{3}{x^2} \left(\cosh x-\frac{\sinh x}{x}\right)\,=\  3\ \sum_0^\infty \ x^{2n}\, \left[\,\frac{1}{(2n+2)!}-\frac{1}{(2n+3)!}\,\right]\ .
\ee
\vspace{1mm}

The expansion  (\ref{expphi0-2}) of $\Phi(x) = \langle\,\frac{\sinh mr}{mr}\,\rangle$ allows for its analytic continuation 
as an holomorphic function of $x$. The change $m\, \to\, \mp\ ik$, i.e., $x^2 \to -\,x^2$  with now $x=kR$, leads to define its dual as 
\be
\label{dual0}
\ba{ccl}
\tilde \Phi(x)=\Phi(ix) =\, \dis \langle \ \frac{\sin kr}{kr}\ \rangle &=&\dis \sum_0^\infty\ (-1)^n\ \frac{x^{2n}}{(2n+1)!}\ \frac{\langle \,r^{2n}\,\rangle}{R^{2n}} 
\vspace{2mm}\\ 
&=& \dis
 1\, - \,x^2\    \frac{\langle\, r^2\,\rangle}{6 \,R^2} +\, x^4\   \frac{\langle\, r^4\,\rangle}{120\,R^4} -\,x^6\ \frac{\langle \,r^6 \,\rangle}{5\,040\,R^6}  
+ \, ...\ .
\ea 
\ee
\vspace{1mm}

\noindent
The expression obtained,
\be
f(k)\,=\ \tilde\Phi(kR)\,=\,\Phi(ikR)\ =
\dis \langle\ \frac{\sin kr}{kr}\ \rangle\,=\,\dis \int_0^R \frac{\rho(r)}{\rho_0} \ \frac{\sin kr}{kr}\ \frac{3r^2 dr}{R^3}\,,
\ee
also written with  a density distribution normalized to unity as
\be
f(k)\,=\,\langle \ e^{i \vec k.\vec r}\ \rangle = \int_0^\infty \!\rho(r)\,2\pi r^2 dr \int_{-1}^1 e^{ikr \cos\theta}\  d(\cos\theta)\,=\, \dis \int_0^\infty 4\pi \rho(r)\ \frac{ \sin kr}{kr}\ r^2dr\ ,
\ee
is  identified as the form factor of the spherically symmetric body considered \cite{pirenne}.
For an homogeneous body it is simply $\tilde\phi(x)=\phi(ix) = \frac{3}{x^2}\,(\frac{\sin x}{x}-\cos x)$, with $x=kR$.

\section{\boldmath Expressing $\,\Phi(x)\,$ from an effective density $\,\bar\rho(x)$}
\label{secrhobar}

An homogeneous sphere has a moment of inertia $\frac25 \,MR^2$ corresponding to $\langle\, r^2\,\rangle = \frac35\,R^2$. 
For the Earth it has a smaller value, usually taken as $I \simeq .3308\,MR^2$  \cite{prem,nasa}, 
even if other values, from .3307 to about .332 $MR^2$, have been considered \cite{denis,ken}. 
This value, close to $\frac13\,MR^2$, corresponds to 
\be
\langle \,r^2\,\rangle\,=\,\frac{3}{2}\ \frac{I}{MR^2}\,\simeq\, .4962 \ R^2\ .
\ee
The Earth density $\rho(r)$ is larger within the core, which results in somewhat smaller values of $\langle \,r^{2n}\,\rangle$, 
and globally of $\Phi(x)$, as compared to $\phi(x)$.

\vspace {2mm}

We can compare, more generally,  the form factor $\Phi(x)$ with the corresponding factor $\phi(x)$ for an homogeneous sphere. 
This leads us to introduce, for each value of the range $\lambda$, an effective average density $\bar\rho(x\!=\!R/\lambda)$, 
such that the inhomogeneous sphere of radius $R$ generates the same outside potential ${\cal V}_{\rm out }(r)$ in (\ref{calV}) 
as an homogeneous sphere with the constant density $\bar\rho(x)$. This  one  is thus defined by 
\be
\label{defrho}
\framebox [4cm]{\rule[-.5cm]{0cm}{1.25cm} $ \dis
 \frac{\bar\rho(x)}{\rho_0} \ =\ \frac{\Phi(x)}{\phi(x)}\ ,
 $}
\ee
\vspace{-5mm}

\noindent
or, more explicitly, as

\vspace{-7mm}

\be
\label{rhobar0}
\bar\rho(x)\ =\ \rho_0\ \,\frac{\Phi(x)}{\phi(x)} \,=\ \frac{ \hbox{$\displaystyle\int_0^R \rho(r)  \,\sinh mr\ rdr $}}{ \hbox{\large $\displaystyle \int_0^R \,\sinh mr\ r dr$}}
\ .
\ee

\noindent
This may also be expressed in terms of the bilateral Laplace transform of $\,r \rho(r)$ (now considered as an odd function of $r$), as
\vspace{-5mm}

\be
\displaystyle \bar\rho(x)\,=\ \,\frac{ \hbox{$\displaystyle\int_{-R}^R r\,\rho(r)\  e^{mr}\ dr$}}{ \hbox{\large $\displaystyle \int_{-R}^R r\,e^{mr}\ dr$}}\ \,=\ \, \frac{ \hbox{$\displaystyle\int_{-R}^R r\,\rho(r)\  e^{-m(R-r)}\ dr$}}{ \hbox{\large $\displaystyle \int_{-R}^R r\,e^{-m(R-r)}\ dr$}}\ \,.
\ee

\vspace{.5mm}

$\bar\rho(x)$ may be viewed as an average of the density $\rho(r)$, weighted proportionally to $({\sinh mr}/{mr})$ 
to take into account  the finite range of the interaction considered. This favors the larger values of $r$, ultimately approaching $R$, 
when $m$ increases so that the range $\lambda = 1/m$ gets smaller.
The ratio $\bar\rho(x)/\rho_0$ can be shown to be $\leq 1$, and decreasing with increasing $x$ i.e. decreasing $\lambda$,  for a density $\rho(r)$ decreasing with $r$ (resp. $\geq 1$ and increasing with $r$,  for a density increasing with $r$).
This effective $\bar \rho(x)$ decreases regularly with decreasing $\lambda$'s, from the average density $\rho_0$ at large $\lambda \gg R$, 
down to the density at the periphery of the sphere, or some average of it, for the smaller values of $\lambda$.

\begin{figure}
\caption{\ For an interaction of range $\lambda $ the Earth generates the same outside Yukawa potential as an homogeneous sphere of density $\bar\rho (x)= \rho_0\,\Phi(x)/\phi(x)$, function of $x=mR=R/\lambda$.
It decreases regularly from about $\rho_0\simeq 5.51$ g/cm$^3$ for $\lambda $ larger than the Earth radius, down to $\simle$ 3.5 g/cm$^3$ for $\lambda \simle $ 300 km.
$\bar\rho (x)$, given by eqs.\,(\ref{defrho}-\ref{limrho}), is then representative of an average density around a depth $d= R-r \approx \lambda$, and is evaluated in Sec.\,\ref{sec:5spheres} in a simple 5-shell model. For $x\simeq 3$ i.e. $\lambda\simeq 2\,100$ km, $\bar\rho\simeq $ 4.92 g/cm$^3$ is close to the inner mantle density, while for $x\simeq 20$ i.e.~$\lambda\simeq 320 $ km, $\bar\rho\simeq$ 3.57 g/cm$^3$ is close to the outer mantle density (cf. Table\,\ref{phiPhirho}).
\label{rho}}
\vspace{5mm}
\includegraphics[width=10cm,height=7cm]{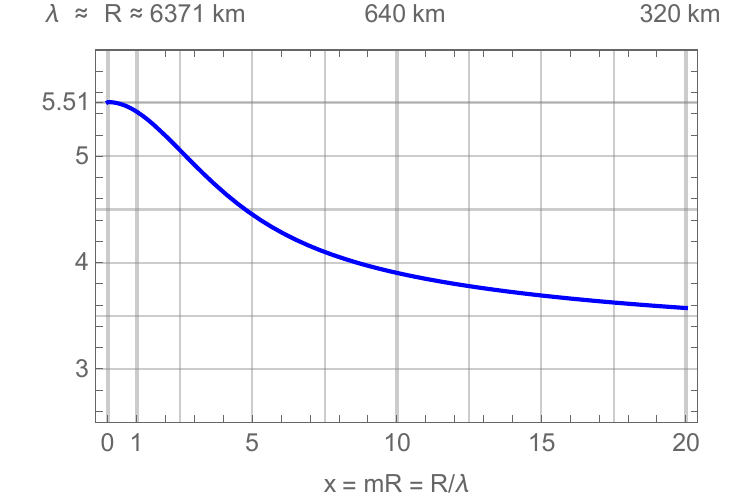}
\vspace{1mm}
\end{figure}

\vspace{2mm}

For $x$ large enough (typically $\simge 5$),  $\bar\rho(x)$ may in general be approximated by
\be
\label{rhobar200}
\dis \bar\rho(x)\ 
\simeq \  
\frac{\hbox{$\displaystyle\int_0^R\  \rho(r)\  \hbox{\large$\displaystyle\normalsize \frac{e^{-d/\lambda}}{r}$}\
 r^2 dr$}}{ \hbox{\large $\displaystyle \int_0^R \ \frac{e^{-d/\lambda}}{r}\ \,r^2 dr$}}
 \ ,
\ee
\vspace{-2mm}

\noindent
$\rho(r)$ 
being simply weighted proportionally to  ${e^{-d/\lambda}}/{r}$, exponentially decreasing with the depth $d=R-r$ below surface.
$\bar\rho(x)$ involves mainly the external shells of the sphere, within a depth $\simle 2\lambda$ below surface,
ultimately decreasing, for small $\lambda$'s,  down to the relevant 
\vspace{1mm}
external density $\rho_e$,
\vspace{-7mm}

\be
\label{limrho}
\frac{\Phi(x)}{\phi(x)}\,=\,\frac{\bar\rho(x)}{\rho_0}\ \to\ \frac{\rho_e}{\rho_0}\,.
\ee

\vspace{1mm}

This one may be taken here as the crust density, close to $\rho_0/2\simeq 2.755$ g/cm$^3$, for small $\lambda$'s of a few tens down to $\approx 20$ km, below which the experiment loses its sensitivity given the mean satellite altitude of  $\simeq $ 710 km. (We can thus ignore the oceanic contribution to $\rho(r)$, that would lead to a lower $\rho_e$.)
\,This provides a reduction factor ${\rho_e}/{\rho_0}$ $\approx \frac12 $
 in the short-range effects of a new force as compared to the homo\-geneous case,
responsible for {\em an increase by $\,\simeq \sqrt 2\,$ of the upper limits on its couplings}.

 \vspace{2mm}
 
The new force is proportional to $\Phi(x) (1+mr)\,e^{-mr}/r^2$, i.e., to
$\Phi(x) (1+mr)\,e^{-mr}$ as compared to the massless case. For small $\lambda$'s this reads
\be
\Phi(x)\,(1+mr)\,e^{-mr}\, \simeq \,\frac{3\, (x-1)}{2x^3}\ (1+mr)\  e^{-m(r-R)}\ \frac{ \bar \rho(x)}{\rho_0} 
\,\simeq \ \frac{3\,\lambda}{2R}\  \frac{\rho_e}{\rho_0} \ \frac{r}{R}\ e^{-mz}\,.
\ee
Near the surface, i.e., at low $z$ compared to $\lambda$, this reduces approximately   to 
{$\,\dis \frac{3\,\lambda}{2R}\,\frac{\rho_e}{\rho_0} $}
\cite{fa860}.

\vspace{3mm}

\section{\boldmath Two ways of viewing a multishell model}

\subsection{\boldmath Expressions of $\Phi(x)$}

Let us approximate the sphere as a series of $n$ homogeneous hollow spheres of radii $R_i$ from $R_1$ to $R_n=R$,  
each one with a density $\rho_i$ varying from $\rho_1=\rho_c$ at the center to $\rho_n=\rho_e$ at the periphery.
The contribution to $\Phi(x)$
of a full homogeneous sphere of radius $R_i$ is $ [\rho_iR_i^3/(\rho_0 R^3)]\, \varphi(x_i)$, with $x_i= R_i/\lambda= x\,R_i/R$. \
Summing on the contributions of the $n$ hollow spheres one gets
\be
\label{calPhi50}
\Phi(x) \,=\,\sum_1^n\ \frac{\rho_i}{\rho_0}\ 
\left[ \,\frac{R_i^3}{R^3}\  \phi(x_i)\ - \frac{R_{i-1}^3}{R^3}\ \phi(x_{i-1})\,\right]\ ,
\ee
with an inside radius $R_0=0$ for the first sphere of radius $R_1$.

\vspace{3mm}

One may  view the same situation as a superposition of $n$ full homogeneous spheres of charges $q_i$  
(or masses $m_i$, when $\rho(r)$ is viewed as a mass density) and radius $R_i$ from $R_1$  to $R_{n}= R$, 
with densities equal to the excess density $\Delta\rho_i =\rho_i-\rho_{i+1}$ (or $\Delta\rho_n =\rho_e$ for the last sphere).
With $\Phi(x)$ providing the external potential as
\be 
{\cal V}_{\rm out}(r)\, = \ Q\ \Phi(x)\ \frac{e^{-mr}}{4\pi r} \,=\, \sum_i\ q_i\ \phi(x_i)\ \frac{e^{-mr}}{4\pi r} \,,
\ee
we get the equivalent expression
\be
\label{calPhi60}
\framebox [12.4cm]{\rule[-.45cm]{0cm}{1.25cm} $ \dis
\Phi(x) \,=\, \sum_1^n\ \frac{q_i}{Q}\ \phi(x_i)\,=\,\sum_1^{n-1}\ \frac{\rho_i-\rho_{i+1}}{\rho_0}\ \frac{R_i^3}{R^3} \ \phi(x_i) \,+\,\frac{\rho_e}{\rho_0}
\ \,\phi(x)\ ,
$}
\ee
with the normalization condition
\be
\label{norm}
\sum_1^n \ \frac{q_i}{Q}\ =\  \sum_1^{n-1}\ \frac{\rho_i-\rho_{i+1}}{\rho_0}\ \frac{R_i^3}{R^3}  \,+\,\frac{\rho_e}{\rho_0}\,=\,1\ .
\ee
This one expresses that for $\lambda = \infty$ so that $x_i= x=0$, all $\phi$ and $\Phi$ in (\ref{calPhi60}) are equal to 1.
\linebreak
Expression (\ref{calPhi60}) of $\Phi(x)$ is a simple rewriting of (\ref{calPhi50}) by reordering its terms.

\subsection{\boldmath $\Phi(x)$ in the continuum limit}

In the continuum limit, the discrete sums  (\ref{calPhi50},\ref{calPhi60}) provide the two integral expressions 
\be
\label{intint}
\framebox [14.2cm]{\rule[-.5cm]{0cm}{1.3cm} $ \dis
\Phi(x)\, = \int_0^R\frac{\rho(r)}{\rho_0}\ d \left[\,\frac{r^3}{R^3}\ \phi(r/\lambda)\,\right] = \,\int_0^R
\frac{-\,d\rho(r)}{\rho_0} \ \frac{r^3}{R^3}\ \phi(r/\lambda)\, + \,\frac{\rho_e}{\rho_0}
\ \phi(x)\ ,
$}
\ee
which are equal, thanks to an integration by parts. The normalization condition (\ref{norm}), which reads
\be
\label{norm2}
\int\,\frac{dq}{Q}\, =\int \,\frac{dm}{M}\,=\,  \int_0^R \,\frac{\rho(r)}{\rho_0} \ \frac{3r^2dr}{R^3} \,=\,
\int_0^R\, \frac{-\,d\rho(r)}{\rho_0}\ \frac{r^3}{R^3}  \ +\ \frac{\rho_e}{\rho_0}\,=\,1\ ,
\ee
expresses that the total new charge (or mass) of the sphere may be decomposed as the one 
of an homogeneous sphere of radius $R$ and density $\rho(R)=\rho_e$, 
\vspace{-.5mm}
plus the contributions of all the homogeneous full spheres of radii $r$, with the infinitesimal densities $-\,\frac{d\rho(r)}{dr}\,dr$\,. 
The second expression in (\ref{intint}) explicitates how $\Phi(x)$ depends 
\vspace{-.5mm}
both of the external density $\rho_e$ 
and on the variations of the relative density inside the sphere, 
described by $\frac{-\,1}{\rho_0}
\, \frac{d\rho(r)}{dr}$.
\vspace{3mm}

With 
\be
d \left[\,\frac{r^3}{R^3}\ \phi(r/\lambda)\,\right]\ =\ \frac{3}{m^3R^3}\ d \,(mr\cosh mr- \sinh mr)\ =\ \frac{\sinh mr}{mr}\  \frac{3r^2dr}{R^3}\,,
\ee
the first expression of $\Phi(x)$ in (\ref{intint}) reads
\be
\label{Phihol}
\Phi(x) \,=\,\dis \int_0^R\,\frac{\rho(r)}{\rho_0}\  \frac{\sinh mr}{mr}\ 
\frac{3\,r^2dr }{R^3}\ = \ \langle \ \frac{\sinh mr}{mr}\ \rangle\ .
\ee
This provides another way to eq.\,(\ref{Phi000}), again obtained as an integral over the contributions 
of the infinitesimally small thin shells of radius $r$ from 0 to $R$. 
\vspace{3mm}

The second expression in (\ref{intint}) may be explicitated as
\be
\label{Phirhoe}
\Phi(x)=\displaystyle  \int_0^R \!
\frac{1}{\rho_0}\,\frac{-\,d\rho(r)}{\ \ dr} \ dr \ \,\frac{3\left(mr\cosh mr-\sinh mr\right)}{(mR)^3}\, +\,\frac{\rho_e}{\rho_0}\  
\frac{3\left(x \cosh  x-\sinh x\right)}{x^3} \ .
\ee
One can also reinclude the last term within the integral, to write
\be
\label{Phirhoe2}
\Phi(x)=\displaystyle  \int_0^\infty \!
\frac{1}{\rho_0}\,\frac{-\,d\rho(r)}{\ \ dr} \ dr \ \,\frac{3\left(mr\cosh mr-\sinh mr\right)}{(mR)^3} \ ,
\ee
again leading back to eq.\,(\ref{Phihol}).
The last term proportional to $\rho_e$ in (\ref{Phirhoe}) is recovered from 
$\,\frac{-\,d\rho(r)}{\ \ dr}=\rho_e\,\delta(r-R) + \, ...\, $.
Eq.\,(\ref{Phirhoe2}) also provides back  $\phi(x)$ for an homogeneous sphere from the discontinuity of $\rho(r)$ at its surface, as
\be
\phi(x)=\int_0^\infty\, \delta(r-R) \ dr  \ \,\frac{3\,(mr\cosh mr-\sinh mr)}{(mR)^3}\,=\ \frac{ 3\,(x\,\cosh x-\sinh x)}{x^3}\ .
\ee

\subsection{\boldmath $\Phi(x)$\, from the evaluation of $\langle \,r^{2n}\,\rangle$ in a multishell model}

Let us consider again the expression (\ref{calPhi60}) of $\Phi(x)$ for a $p$-shell model of the sphere, 
viewed as a superposition of $p$ homogeneous full spheres of radii $R_i$ and charges $q_i$ (or masses $m_i$), 
or from expression (\ref{intint}) obtained in the continuum limit. It reads,  replacing the charge ratios $q_i/Q$ 
by the corresponding mass ratios $m_i/M$,
\be
\label{shell}
\Phi(x)\,=\, \sum _{i=1}^p\ \frac{m_i}{M}\ \phi(R_i/\lambda)\ \to\ \int\,\frac{dm}{M}\ \phi(r/\lambda)\ ,
\ee
where $\phi(R_i/\lambda)$ may be expanded as in (\ref{expphi7}), according to 
\be
\phi(R_i/\lambda)\,=\,\sum_{n=0}^\infty \ \frac{x^{2n}}{(2n+1)!}\ \frac{R_i^{2n}}{R^{2n}}\ \frac{3}{2n+3}\ .
\ee
With the moments of  $\rho(r)$ given by
\be
\label{expmom}
\langle\, r^{2n}\,\rangle\ = \, \frac{3}{2n+3}\  \sum_{i=1}^p\ \frac{m_i}{M}\ R_i^{2n}\ \to\  \frac{3}{2n+3}\  \int\,\frac{dm}{M}\ r^{2n}\,,
\ee
\vspace{-5mm}

\noindent
this reads
\be
\Phi(x)=\int \frac{dm}{M}\, \phi(r/\lambda)=\,\sum_{n=0}^\infty  \frac{x^{2n}}{(2n+1)!}\, 
\int\frac{dm}{M} \, \frac{r^{2n}}{R^{2n}}\, \frac{3}{2n+3}
\, =  \,\sum_{n=0}^\infty \, \frac{x^{2n}}{(2n+1)!}\, \frac{\langle\, r^{2n}\,\rangle}{R^{2n}}\, ,
\ee

\noindent
which reconstitutes the expansion (\ref{expphi0-2}),
\vspace{-.5mm}
absolutely convergent for all $x$, and identified as
\hbox{\Large $\langle \, \frac{\sinh mr}{mr}\, \rangle$}\,.

\vspace{4mm}

This provides a fourth way to obtain the general expression of $\Phi(x)$, in addition to summing the contributions of thin shells 
and solving the Poisson-like equation for the potential as in (\ref{poiss}-\ref{Phi000}), 
decomposing the sphere into  small hollow spheres of radius $r$ and thickness $dr$ as in (\ref{calPhi50}-\ref{Phihol}), 
or viewing it as a superposition of an homogeneous sphere of radius $R$ and density $\rho_e$ 
with a set of smaller full spheres of radius $r$ from 0 to $R$  as in (\ref{calPhi60}-\ref{Phirhoe}).

\vspace{3mm}

To go further we need to specify the density distribution $\rho(r)$.
We have, from the Earth moment of inertia $I= \frac{2}{3}\,M\,\langle \,r^2\,\rangle \simeq .3308\,MR^2 $,
$
\langle \,r^2\,\rangle\,\simeq \,.4962\ R^2,
$
\,so that
\be
\Phi(x)\,\simeq \,1 + \,.0827\ x^2 \,+\, ...\ .
\ee
This is slightly below $\phi(x)= 1 + \frac{x^2}{10}+ ...\, $ for an homogeneous sphere,  in agreement with the fact that 
\vspace{-6mm}

\be
\label{exprhobar}
\frac{\bar\rho(x)}{\rho_0}\,=\,\frac{\Phi(x)}{\phi(x)}\,\simeq\,1\,-\,.0173 \ x^2 \,+\,...\ 
\ee 
is a decreasing function of $x$, as illustrated in Fig.\,\ref{rho}. 
\,For $\,\lambda =R\,$ i.e. $x=1$ this leads to $\bar\rho(1)/\rho_0\simeq .983$, close to the .9835 shown later in Table \ref{phiPhirho} 
(in agreement with the small $x$ expansion $\,\bar\rho(x)/\rho_0\,\simeq 1-.0173 \  x^2 + 8.65 \times 10^{-4}\ x^4 + ...\,$).

\vspace{1mm}

\section{The Earth as a superposition of 5 homogeneous full spheres}
\label{sec:5spheres}

We now consider a simplified model of the Earth, taken as composed of five  sectors of external radii $R_i$ and densities $\rho_i$. 
We then view it as a superposition of five homogeneous full spheres of radii $R_i$ increasing from $R_1=R_{ic}$ to $R_5=R$, 
each one having for density the excess density 
\be
\Delta \rho_i = \rho_i-\rho_{i+1}\,,
\ee
the last one with density
\be
\Delta \rho_5 = \rho_5 = \rho_{\rm crust}= \rho_0/2 = 2.755 \ \hbox{g/cm}^3 \,.
\ee
We take $R_5=R= 6\,371$ km, an outer mantle radius $R_4=R_{om}= $ $6\,341$ km (for a crust of thickness $\simeq $ 30 km), 
an inner mantle radius $R_3 = R_{im} = 5\,701$ km, and outer and inner core radii $R_2=R_{oc}$ = 3\,480 km 
and $R_1=R_{ ic}= 1221$ km  \cite{prem}.

\begin{table}
\caption{
A 5-shell model of the Earth, with five sectors of densities $\rho_i$ between radii $R_{i-1}$ and $R_i$. It is then viewed as a superposition of five full spheres of radii $R_i$, masses $m_i= y_i \,M$ and densities $\Delta \rho_i = \rho_i-\rho_{i+1}$, the fifth one, of radius $R$, with density $\Delta\rho_5 = \rho_{\rm crust}=\rho_0/2=2.755 $ g/cm$^3$.
The outer mantle density, and the difference between inner and outer core densities, 
are taken as 3.5 and 1.8 g/cm$^3$.
The mass and moment of inertia of the Earth are reproduced with
$\,\sum_i y_i = 1$ and $\,\sum_i y_i\,{R_i^2}/{R^2}= I/(\frac25\,MR^2)\simeq.827 $. 
\vspace{-.3mm}
}
\label{5shell}
\vspace{4mm}
\normalsize
$\ba{|c||c|c|c|c|c|}
\hline
&&&&& \\ [-3mm]
&  R_5= R &R_4 = R_{om}& R_3= R_{im}& R_2= R_{oc}&R_1 = R_{ic}
\\ [1mm]
& \  = 6\,371\ \hbox{km}\ & \ = 6\,341\ \hbox{km} &  \ = 5\,701\ \hbox{km} \  & \ = 3\,480\ \hbox{km} \  & \ = 1\,221 \ \hbox{km} 
\ \\ [2mm]
\hline \hline &&&&& \\ [-3.5mm]
{R_i^2}/{R^2}
&1 & . 9906 & . 8007&.2984& .0367
\\ [1.3mm]
\hline &&&&& \\ [-4mm]
 & \rho_{\rm crust}&  \  \,\rho_{om}-\rho_{\rm crust} \ \,  &\  \rho_{im}-\rho_{om}\ & \rho_{oc}-\rho_{im} &\rho_{ic}-\rho_{oc}
\\
&&&&& \\ [-4mm]
\ba{c}
\vspace{-12mm}\\
\Delta \rho_i = \rho_i\!-\!\rho_{i+1}\  \ (\hbox{g/cm}^3)
\ea & = \,2.755& =\,  .745 &= \, 1.319 &= \, 6.522 & =\, 1.8
\\ [2mm] \hline 
&&&&& \\ [-3.5mm]
\rho_i\ (\hbox{g/cm}^3) & 2.755& 3.5 &4.82 &11.34&13.14
\\ [1.5mm]
\hline  &&&&& \\ [-4mm]

\ \ba{c}
\hbox{mass fraction:}\vspace{.5mm}\\
y_i= \dis  \frac{m_i}{M}=\,\frac{\Delta \rho_i}{\rho_0}\, \frac{R_i^3}{R^3}
\vspace{-1mm}\\
\ea
\ 
 &.5 & .1333 & .1715 & .1929 & .0023
\\ [6.5mm]
\hline
&&&&& \\ [-3.5mm]
\ \ y_i\ R_i^2/R^2\ \ \, & .5& .1320& .1373 & .0576& .0001
\\ [1.5mm]
\hline
\ea $
\vspace{3mm}
\end{table}

\vspace{2mm}

We include for completeness a distinction between inner and outer cores, even if it is inessential in view of the small effect 
of an increased inner core density, 
over .7 \% of the Earth volume, corresponding to $\simle 3 \times 10^{-3}$ of its mass  (or $\simeq 2.3 \times 10^{-3}$,
with the excess density 1.8 g/cm$^3$ chosen here). 
In fact the structure of the density distribution within the core has only very small or negligible effects for the smaller ranges $\lambda\,$. 
For larger ones (i.e. small or moderate $x=R/\lambda$), once $\langle \,r^2\,\rangle$ is fixed to .4962 $R^2$ 
by the moment of inertia $I\simeq .3308\ MR^2$, the effects of an increased inner core density mainly occur 
through  the adjustment of the $\,x^4\, \langle \,r^4\,\rangle/(120\ R^4)$ term in expression (\ref{expphi0-2}) of $\Phi(x)$, and are very small.
We shall consider here that the excess of the inner over outer core densities is $\rho_1 = \rho_{ic}-\rho_{oc} \simeq 1.8$ g/cm$^3$, 
the results being largely insensitive to this choice.
\vspace{2mm}

We take the outer mantle density as  $\rho_{om} = 3.5$ g/cm$^3$, and adjust the inner mantle and outer core densities 
to reproduce the Earth mass and moment of inertia $I\simeq .3308 \ MR^2$.
The proportions of the total mass $M$ in each one of the five full-sphere contributions are denoted as
\vspace{-7mm}

\be
\label{y}
y_i=\frac{m_i}{M}= \frac{\Delta \rho_i}{\rho_0}\ \frac{R_i^3}{R^3}\,,\ \ \ \hbox{with}\ \ \sum_i\ y_i= 1\,,
\ee
and
$\,I= \sum _i \, \frac25\,   m_i  R_i^2=
\sum_i \, y_i \, \frac{R_i^2}{R^2} \times  \frac25\, MR^2 \simeq .3308 \,MR^2,
$
\,so that
\be
\label{y2}
\sum_i \ y_i \  \frac{R_i^2}{R^2}\,\simeq .8270\ .
\ee
The mass fractions $y_5= .5, \ y_4\simeq  .1333$ and $y_1\simeq .0023$ are fixed by the chosen values of the crust and outer mantle densities 
(for $y_5$ and $y_4$), and excess of the inner over outer core densities (for $y_1$). The two remaining fractions $y_3$ and $y_2$, 
satisfying $y_3+y_2\simeq .3644$, are adjusted to $y_3\simeq .1715$ and $y_2\simeq .1929$ so as to reproduce the Earth moment of inertia. 
Altogether we get, as shown in Table \ref{5shell},
\be
\left\{\ \ba{crrrrrcc}
y_5 \simeq .5\,,&\, y_4 \ \simeq\, .1333\,,  &\, y_3\ \simeq \,.1715\,, &\,y_2 \ \simeq\, .1929\,,&\, y_1 \ \simeq \,.0023\,,
\vspace{2mm}\\
 y_5 \simeq .5\,, &\, y_4 \, \frac{R_{om}^2}{R^2} \simeq \,.1320\,,  &\,y_3\, \frac{R_{im}^2}{R^2}\simeq \,.1373\,, &\,y_2\, \frac{R_{oc}^2}{R^2}\simeq \,.0576\,, &\, y_1\, \frac{R_{ic}^2}{R^2}\simeq\, .0001\,. \ea\right.  
 \ee
\vspace{2mm}

This fixes the density differences
\be
\left\{\ 
\ba{ccccccc}
\Delta \rho_3 \!&=&\! \rho_{im}- \rho_{om} &=& y_3\ \frac{R^3\ }{R_{im}^3}\ \rho_0 &\simeq& 1.319\ \,\hbox{g/cm}^3,
\vspace{2mm}\\
\Delta \rho_2  \!&=&\! \rho_{oc}- \rho_{im}  &=& y_2\ \frac{R^3\ }{R_{oc}^3}\ \rho_0  &\simeq&  6.522\ \,\hbox{g/cm}^3,
\ea \right.
\ee
leading to
\be
\label{rhoim}
\rho_{im}\,\simeq 4.819\ \hbox{g/cm}^3,\ \ \rho_{oc}\,\simeq 11.341\ \hbox{g/cm}^3, \ \ \rho_{ic}\,\simeq 13.141\ \hbox{g/cm}^3.
\ee
Within this simplified 5-shell model 
one can estimate, from eq.\,(\ref{calPhi60}), 
\be
\label{Phi5}
 \framebox [14.4cm]{\rule[-.8cm]{0cm}{1.8cm} $ \dis
\ba{cccccl}
\Phi(x)\,=\,\dis  \sum _i \,y_i\ \phi(R_i/\lambda) \!&\simeq  \ .5\ \phi(x) \!&+&\! .1333\ \phi(.9953\ x) \!&+&\! .1715\ \phi(.8948 \ x)\ \,
\vspace{-1mm}\\
& \hspace{10mm}\dis    \!&+&\!  .1929\ \phi(.5462\ x) \!&+&\! .0023\  \phi(.1916\ x)\ .
\ea
$}
\ee
\vspace{2mm}

\begin{figure}
\caption{\ The hyperbolic form factors $\phi(x)=3\,(x\cosh x-\sinh x)/x^3$  for an homogeneous sphere (in orange), 
and $\Phi(x)$, as approximated in  a 5-shell model of the Earth (in blue), as functions of $x=mR=R/\lambda$. 
Their ratio $\Phi(x)/\phi(x) = \bar \rho(x)/\rho_0$ defines the effective density $\bar\rho(x)$ represented earlier in Fig.\,\ref{rho}.
\label{phiphi}
\vspace{3mm}}
\includegraphics[width=10cm,height=7cm]{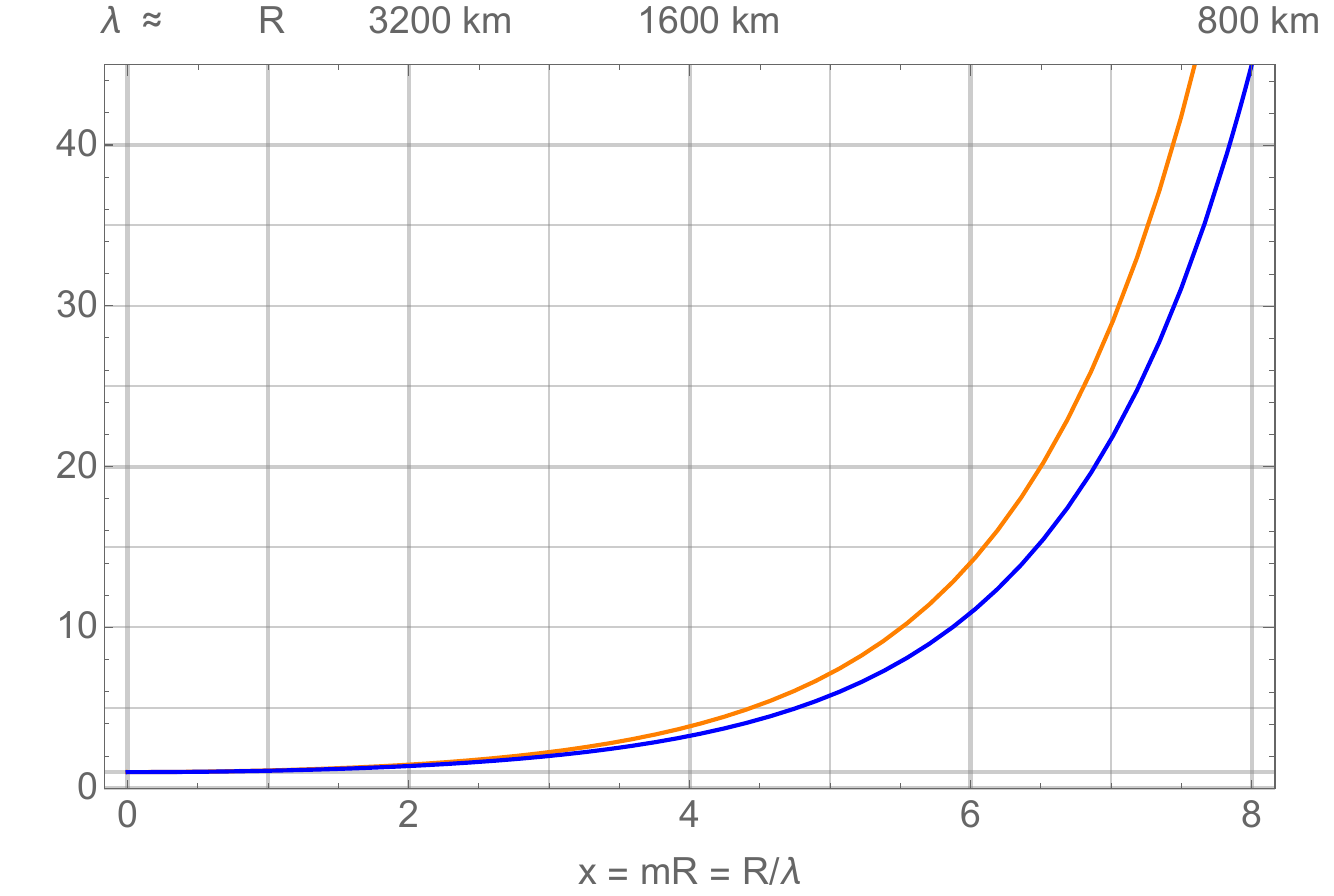}
\end{figure}

The resulting values of $\Phi(x)$, and $\bar\rho(x)/\rho_0$, are given in Table \ref{phiPhirho}. 
They are represented in Fig.\,\ref{rho} for $\bar\rho(x)$, and Fig.\,\ref{phiphi} for $\phi(x)$ and $\Phi(x)$.
As an illustration for $x=3$ (i.e. $\lambda\simeq 2\,124$ km or $m\simeq .929\times 10^{-13} $ eV/$c^2$), $\Phi(3)\simeq 2.0021$ 
as compared to $\phi(3) \simeq 2.2428$ for an homogeneous sphere, corresponding to $\bar\rho(3)/\rho_0 \simeq .893$, or to an effective density 
\be
\bar\rho(3) \,= \,\frac{\Phi(3)}{\phi(3)}\ \rho_0\,\simeq\, .893\ \rho_0\,\simeq  \,4.92 \ \hbox{g/cm}^3\, .
\ee
This is not far from the inner mantle density $\rho_{im} \simeq 4.82 $ g/cm$^3$ obtained in eq.\,(\ref{rhoim}), 
and in agreement with expression (\ref{rhobar200}) of $\bar\rho(x)$. 
For $x=20$ (i.e. $\lambda\simeq 319$ km or $m\simeq 6.2\times 10^{-13} $ eV/$c^2$), $\Phi(20) \simeq 1.121\times 10^6$, 
to be compared with $\phi(20)\simeq 1.728\times 10^6$, leading to
\be
\bar\rho(20) \,= \,\frac{\Phi(20)}{\phi(20)}\ \rho_0\,\simeq\, .648\ \rho_0\,\simeq  \,3.57 \ \hbox{g/cm}^3\, ,
\ee
not far from the chosen value of the outer mantle density $\rho_{om} = 3.5 $ g/cm$^3$ (cf. Fig.\,\ref{rho}).

\begin{table}
\caption{$\phi(x=R/\lambda)$ for an homogeneous sphere, and $\Phi(x)= \sum_i \frac{m_i}{M}\ \phi(R_i/\lambda)$ 
 in eq.\,(\ref{Phi5}) for the 5-shell model of Table \ref{5shell}. \,Their ratio defines the effective density $\,\bar\rho(x)= \rho_0 \,\Phi(x)/\phi(x)$, \hbox{decreasing from $\simeq \rho_0$ for large $\lambda\simge R$}, down to somewhat above $\rho_{\rm crust}\simeq \frac12\,\rho_0$, for shorter $\lambda$'s down to a few tens of km. Cf. Figs.\,\ref{rho} and \ref{phiphi}.\ 
\vspace{-.3mm}
\\ [-2mm]  \\
\hspace*{5mm} 
}
\label{phiPhirho}
\vspace{-1mm}
$
\ba{|c|c||c|c|c|}
\hline
&&&&\\ [-2.5mm]
\  x =mR= R/\lambda\  &\ \  \lambda \ \hbox{(km)}\ \ 
 \ &\ \ \ \  \phi(x)\ \ \ \ &\ \ \ \  \Phi(x)\ \ \ \  & \ \ \displaystyle \frac{\bar \rho(x)}{\rho_0}= \frac{\Phi(x)}{\phi(x)}\ \
\\ [3mm]
\hline\ghost{}\hline
&&&&\\ [-4mm]
 .5 &12\,742 \ \ & 1.0252 &  1.0208 &.996 
\\ [.7mm]
\hline
&&&&\\ [-3.8mm]
1 & 6\,371 &1.1036 & 1.0855 &  .984 
\\
&&&&\\ [-4.5mm]
2 &  3\,185& 1.4616 & 1.3773 &  .942
\\ 
&&&&\\ [-4.5mm]
3 & 2\,124 & 2.2428 &2.0021 &  .893 
\\
&&&&\\ [-4.5mm]
5 & 1\,274 &7.1243 & 5.7563 & .808 
\\ [1mm]
\hline 
&&&&\\ [-3.8mm]
10 & \ \ 637 &\ 297.36 \ & 210.7 & .708
\\ 
&&&&\\ [-4.5mm]
20 & \ \ 319&\ 1.728\times 10^6 \ &\ 1.121\times 10^6\ &.648
\\
&&&&\\ [-4.5mm]
30 & \ \ 212&\ \,1.722\times 10^{10}\, &\ \,1.078\times 10^{10} \,& .626
\\ 
 &&&&\\ [-4.5mm]
50&\ \ 127 &\ \,3.049\times 10^{18}\, &\ \,1.852\times 10^{18} \, & .607
\\ [1mm]
\hline
 &&&&\\ [-3.8mm]
100& \ \ \  63.7 &\ \,3.992\times 10^{39}\, &\ \,2.332 \times 10^{39} \, & .584
\\ 
 &&&&\\ [-4.5mm]
200&\ \ \ 31.9 &\ \ \,2.696\times 10^{82}\ \, &\ \ \,1.490\times 10^{82} \ \, & .553
 \\ [1.2mm]
\hline
\ea
$
\vspace{1mm}
\end{table}

\section{\boldmath Explicit expansion of $\,\Phi(x)$}

In the continuum limit, treating apart the contribution of the last sphere of radius $R$, 
\be
\sum_i\ \frac{m_i}{M} \, \frac{R_{i}^{2n}}{R^{2n}}\ \, \to\ \,\int_0^R\, \frac{-\,d\rho(r)}{\rho_0}\ \frac{r^3}{R^3}\ \frac{r^{2n}}{R^{2n}}\, +\,\frac{\rho(R)}{\rho_0}\,=\, (2n+3) \int_0^R\, \frac{\rho(r)}{\rho_0}\ \frac{r^{2n+2}\, dr}{R^{2n+3}}
\ ,
\ee
which provides 
\be
\label{r2n}
A_{2n}\,=\,\frac{\langle \,r^{2n}\,\rangle}{R^{2n}} =  \,\sum_i\,\frac{m_i}{M}\  \frac{3}{2n+3}\ \frac{R_{i}^{2n}}{R^{2n}}\ \ \ \to\ \ 
\, \int_0^R\, \frac{\rho(r)}{\rho_0}\ \frac{3\,r^{2n+2}\, dr}{R^{2n+3}}\,,
\ee
and 
\be
\Phi(x)\,=\,\sum_0^\infty\ \frac{x^{2n}}{(2n+1)\,!}\ \frac{\langle \,r^{2n}\ \rangle}{R^{2n}}\,=\,\sum_0^\infty\ \frac{x^{2n}}{(2n+1)\,!} \  \int_0^R\ \frac{\rho(r)}{\rho_0}\ \frac{3\,r^{2n+2}\, dr}{R^{2n+3}}\ \,.
\ee

\vspace{3mm}

In the above 5-shell model of the Earth, for which $\Phi(x)$ is given by eq.\,(\ref{Phi5}), 
\be
 \sum_i\,y_i \, \frac{R_{i}^{2n}}{R^{2n}}\, \simeq \, .5 \,+ .1333 \times .9953^{\,2n} + .1715 \times .8948^{\,2n}
+ .1929 \times  .5462^{\,2n} + .0023\times .1916^{\,2n}, 
\ee

\vspace{-3mm}

\noindent
with  $y_i = \dis {m_i}/{M}$, we get  from eq.\,(\ref{r2n}) 
\be
\label{mome}
\left\{ \,
\ba{ccccr}
\langle \,r^{2}\,\rangle\!&=&\, \sum_i\,y_i \  \frac{R_i^{2}}{R^{2}}\,  \times\ \frac{3}{5}\ R^{2} \!&\simeq & .4962\ R^{2}\,,
\vspace{.5mm}\\
&&\hspace{15mm} ...
\vspace{.5mm}\\
\langle \,r^{2n}\,\rangle\!&=&\, \sum_i\,y_i \ \frac{R_i^{2n}}{R^{2n}}\,  \times\, \frac{3}{2n+3}\ R^{2n} \!&=&  A_{2n}\ R^{2n}\,,
\vspace{1mm}\\
&&\hspace{15mm} ...
\ea \right.
\ee
\vspace{-5mm}

\noindent
with
\be
\label{A}
A_0,\,A_2,\,...\,,A_{14},\,... \,\simeq\,1\,,\ .4962\,,\ .3248\,,\ .2409\,,\ .1910\,,\ .1579\,,\ .1343\,,\ .1166\,,\,...\,.
\ee
$A_{2n}$ behaves at large $n$ much like $3/(2n+3)$ times .6333, then .5.

\vspace{3mm}

This provides
\be
\label{devPhi}
\hspace{-20mm}
\Phi(x)\,=\,\dis\sum_0^\infty\ \frac{x^{2n}}{(2n+1)!}\ 
\frac{3}{2n+3}\ \sum_i\,y_i\ \frac{R_i^{2n}}{R^{2n}}
\,=\ \sum_0^\infty\ \frac{A_{2n}}{(2n+1)!}\ \,x^{2n}\ ,
\ee
or explicitly,
in the 5-shell model considered,
\be
\label{devPhi2}
\dis 
\Phi(x)\,\simeq\, 1+ \hbox{\normalsize $\dis \frac{.4962}{6}\ x^2+ \frac{.3248}{120} \ x^4+ \frac{.2409}{5\,040}\ x^6+\frac{.1910}{9\,!}\ x^8+\frac{.1579}{11\, !}\ x^{10}+ \frac{.1343}{13\, !}\ x^{12}+\frac{.1166}{15\, !}\ x^{14}+\,... $}\,,
\ee
i.e.
\vspace{-3mm}

\be
\label{dev}
\framebox [15.5cm]{\rule[-.65cm]{0cm}{1.6cm} $ \dis
\ba{ccl}
\Phi(x)& \simeq&\dis 1\, +\, .0827\ x^2 +\, (2.707 \times 10^{-3})\ x^4+ \,(4.78\times 10^{-5}) \ x^6+ \ (5.26\times 10^{-7}) \ x^8
\vspace{2mm}\\
&& + \ (3.95\times 10^{-9}) \ x^{10}+\,(2.16\times 10^{-11})\ x^{12} 
+\,(8.92 \times 10^{-14})\ x^{14} +
...\ ,
 \ea
$}\,
\ee
as also obtained directly from the expansion of eq.\,(\ref{Phi5}). More specifically we have
\be
\ba{ccc}
\Phi(1,2,3,4,5\,;\,6,7,8,9,10) &\simeq& 1.0855,\ 1.3773,\ 2.0021, \ 3.2510, \ 5.756\,;
\vspace{2mm}\\
&& 10.89, \ 21.70, \
44.96, \ 96.11, \ 210.7\,;
\ea
\ee
which illustrates how $\Phi(x)$ increases rapidly with $x$.
This expansion, absolutely convergent for all $x$, already provides, from its first five terms up to $x^8$ 
(summing up e.g. to $\simeq 10.60$ for $x=6$, not far from $\Phi(6) \simeq 10.89$) a good approximation of $\Phi(x)$ up to $x\simeq 6\,$.

\vspace{2mm}

Eqs.\,(\ref{mome}-\ref{devPhi2}) show how improvements of the Earth model considered would affect, normally in a modest way, the above expansion (\ref{dev}) of $\Phi(x)$.  Any adjustment of the moments as compared to eq.\,(\ref{mome},\ref{A}) would lead to a change in the expression of $\Phi(x)$, by
\be
\delta\,\Phi(x) \,=\,\sum_1^\infty\  \frac{x^{2n}}{(2n+1)!} \ \frac{\delta \,\langle r^{2n}\rangle}{R^{2n}}\ .
\ee

\begin{table}[t]
\caption{
The strength of a finite range force, as compared to the massless or ultralight case, 
 is fixed by  $F(x)=(1+mr)$ $e^{-mr}\, \Phi(x)$,  with $mr \simeq 1.1114\ x$ at $z\!=\! 710$~km. 
 $\Phi(x) = 3\,(x\cosh x-\sinh x)/x^3 \times \bar \rho(x)/\rho_0$
\vspace{-.3mm}
 is given in Table \ref{phiPhirho}.
 \vspace{-.5mm}
The coupling limits in eqs.\,(\ref{lfinal41},\ref{lfinal40}) 
get multiplied by  $G(x)=F(x)^{-1/2}= e^{mr/2}/(\sqrt{1\!+\!mr} \,\sqrt{\Phi(x)}\,)  $, 
 behaving at large~$x$ like $mR\ e^{mz/2}/\sqrt{1+mr}$.
\,For $m= 10^{-13}, \,10^{-12}$ or $10^{-11}$ eV$/c^2$, 
$G(x)\simeq 1.9$, 34, or $10^9$. The resulting coupling limits are shown in Table \ref{limg} and Fig.\,\ref{lim}.
\\ [1mm] 
}
\label{limitg}
\vspace{0mm}
$
\,\ba{|c||c|c|c|c|c|c|}
\hline
&&&&&&\\ [-3mm]
 \ \  m\,(\hbox{eV}/c^2)\ \   &\ \lambda\,(\hbox{km})\  & \, x= R/\lambda \, &\dis  \frac{\bar\rho}{\rho_0}=\frac{\Phi}{\phi}&\ \ \ \ \ \Phi(x)\ \ \ \ \  
 & F(x)\!=\!(1\!+\!mr)\,e^{-mr}\,\Phi(x) 
 &
 \dis \,G(x)\!=\!\frac{e^{mr/2}}{\sqrt{1\!+\!mr}\,\sqrt{\Phi(x)}\,}
\\ [4mm]
\hline
&&&&&&\\ [-3.5mm]
10^{-14}  & 19\,733\, & .323 & .998 &1.0087&.957  & 1.022 
\\ [2mm]
&&&&&\\ [-6.5mm]
2\times 10^{-14}  & \!9\,866\!& .646& .993 & 1.0350 & .867 &1.074
\\ [2mm]
&&&&&\\ [-6.5mm]
5\times 10^{-14}  & \!3\,947\!& 1.614 & .960 &1.2348 & .574 & 1.320  \\ [1mm]
\hline
&&&&&&\\ [-3.5mm]
 10^{-13} & \!1\,973\!& 3.23 &  .882 & 2.217& .281 &1.886
\\ [2mm]
&&&&&\\ [-6.5mm]
2\times 10^{-13}  & 987& 6.46 & .766  &14.8  & 9.27 \times 10^{-2}&3.28
\\ [2mm]
&&&&&\\ [-6.5mm]
 5\times  10^{-13}&395 & 16.14 & .664 & 3.68 \times 10^4& \!\!1.124 \times 10^{-2} &9.43 
\\ [1mm]
\hline
&&&&&&\\ [-3.5mm]
10^{-12}&197& 32.3 &.623 & 9.13 \times 10^{10} & 8.77 \times 10^{-4} &
33.8
\\ [2mm]
&&&&&\\ [-6.5mm]
2\times 10^{-12}&\ \,98.7 & 64.6& .600&2.35 \times 10^{24} & 1.16 \times 10^{-5} & 294
\\ 
&&&&&\\ [-4.5mm]
5\times 10^{-12}&\ \,39.5& 161.4 & .563 &4.14\times 10^{65}& \  \,8.93\times 10^{-11}&1.06\times 10^5
\\ [1mm]
\hline 
&&&&&&\\ [-3.5mm]
10^{-11} &\ \,19.7 & 323 & .530&  \!1.26\!\times \!10^{135}\!& \ \,6.46\times 10^{-19} &1.24  \times 10^9
\\ [1mm]
\hline
\ea\,
$
\vspace{2mm}
\end{table}

\section{\boldmath Limits on the couplings as functions of $m$, or $\lambda$}

The Yukawa potential $\,{\cal V}(r)\! = (Q\ e^{-mr}\!/4\pi r)\, \Phi(x)\,$ leads to a field
${\cal E}(r)=
(Q\ e^{-mr}\!/4\pi r^2) $ $\,(1+mr)\, \Phi(x),
$
and to an E\" otv\"os parameter expressed from (\ref{deltaeps0}) as
\be
\delta\,= \ \mp \ \frac{\alpha}{G_N\,u^2}\ 
\epsilon^2 \, \left(\hbox{\normalsize $\dis \frac{Q}{A_r}$}\right)_\oplus 
\Delta \! \left(\hbox{\normalsize $\dis \frac{Q}{A_r}$}\right)
\,  e^{-mr}\ (1+mr)\ \Phi(x)\ .
\ee
It is multiplied by $F(x) =  e^{-mr}\ (1+mr)\ \Phi(x)$, as compared to the ultralight or massless case. For small $\lambda$ i.e. large $x$ one has
\be
F(x)  
\,\simeq\, \frac{3\, (1-\frac1x)}{2x^2}\, (1+mr)\,  e^{-m(r-R)}\ \frac{ \bar \rho(x)}{\rho_0} 
\,\approx\ \frac{3\lambda}{2R} \, \left(1+\frac{z}{R}\right)\,  e^{-mz}\ \frac{\bar \rho(x)}{\rho_0}\ .
\ee
The resulting limits on the $\epsilon$ parameters, and on the couplings $g=\epsilon \, e$, get increased by the factor

\vspace{-7mm}

\be
G(x)\,=\,F(x)^{-1/2}
\,=\,\frac{e^{mr/2}}{\sqrt {\phi(x)}\ \sqrt {1+mr}}\  \,\sqrt \frac{\rho_0}{\bar\rho(x)}\ ,
\ee 
\vspace{-2.5mm}

\noindent
with
$\phi(x) = \frac{3}{x^2}\left(\cosh x-\frac{\sinh x}{x}\right)$\,.

\vspace{3mm}

The values of $\Phi(x),\, F(x)$ and $G(x)$  are given in  Table \ref{limitg}
\,\footnote{With $m=10^{-12}$ eV/$c^2$ 
corresponding 
\vspace{-.3mm}
to $\lambda\simeq 197.327 $ km and $x\simeq 32.287$,  $\phi(x)\simeq 1.466\times 10^{11}$, $ \Phi(x)\simeq 9.13\times \!10^{10}$, and $ \bar\rho(x)/\rho_0\simeq .623$ as in Tables \ref{phiPhirho},\,\ref{limitg}, 
i.e., an effective density $\bar\rho(x)\simeq 3.43$ g/cm$^3$ and $ \sqrt {{\rho_0}/{\bar\rho(x)}}\simeq$ $1.267$. \,With $(1+mr)\,e^{-mr}\simeq \,9.60\times 10^{-15}$, 
\vspace{-.5mm}
the new force gets multiplied by $F(x)\simeq 8.77$  $ \times \, 10^{-4}$,\, and the coupling limits increase by the factor $G(x)
=F(x)^{-1/2}\simeq 33.8$ as in Table \ref{limitg}, also in agreement with (\ref{gapp}).}.
At large $x$, with $mr = x \, (1\!+ z/R) \simeq 1.1114\ x$ and $mz/2\simeq .0557 \ x$, one gets  the simplified expression
\be
\label{gapp}
G(x)\,\simeq\,\sqrt{\frac{2}{3\,(1-\frac1x )}} \ \,x\ \frac{e^{mz/2}}{\sqrt{1+mr}} \  \sqrt \frac{\rho_0\ }{\bar\rho(x)}
\,\approx\ .77\  \sqrt x\ \,e^{0.0557 \, x} \  \sqrt {\frac{\rho_0\ }{\bar\rho(x)}}\ ,
\ee
where $\sqrt{\rho_0/\bar\rho(x)}$ increases from $\simeq 1$ at small $x$ up to $\simeq \sqrt 2$ at larger values of $x$, 
which provides a good approximation of $G(x)$.
The large factor $\Phi(x)$ is essential for the smaller $\lambda$'s, as  
the effect of the new force does not decrease proportionally to $e^{-mr}$ but only  to $e^{-mz}$
\vspace{-.6mm}
($z\simeq 710 $ km being the mean satellite altitude). 
Ignoring  $\Phi$ would lead to overestimate the limits by a factor $\sqrt\Phi$, 
which is $\,\approx 3\times 10^5$ for $m\simeq 10^{-12}$ eV/$c^2$.

\begin{table}[t]
\caption{\ Upper limits (95 \% CL) on the couplings of a spin-1 or spin-0 particle effectively coupled to $B$, $B\!-\!L$, or $L$. 
They  follow from the expressions and limits on $\delta$  in (\ref{deltaeps0},\ref{limdelta}), leading to the $m=0$ limits  (\ref{lfinal41},\ref{lfinal40}), 
multiplied by $G(x)$ from Table \ref{limitg}. 
For a given sign of $\delta$ the limits on $|g_B|$ and $|g_L|$  are larger than for $|g_{B-L}|$ by about 6 and 1.02, respectively; 
for $\delta <0$ they are larger than for $\delta >0$ by $\simeq 1.2$. 
The limits on $|g_{B-L}|,\, |g_B|$ and $|g_L|$, are approximately proportional to [1, $(6\times\! 1.2)$ and $(1.02\times \!1.2)$] 
and [1.2, 6 and 1.02] in the spin-1 and spin-0 cases, respectively. 
These limits, $\simeq 1.07 $ $\times \,10^{-15},\   7.7 \times 10^{-15}$ and $1.31 \times 10^{-15}$ 
in the ultralight spin-1 case (and $ 1.28 \times \,10^{-15},\   6.4 \times 10^{-15}$ and $1.09 \times 10^{-15}$ for spin-0)  
are represented in Fig.\,\ref{lim} as functions of $m$.
}
\label{limg}
\vspace{4mm}
$
\ba{|r|c||c|c||c|c|}
\hline 
&&&&&\\
&&&&&\\ [-6mm]
 &&
  \hbox{lim}\, \hbox{\boldmath $ |g_B|$}& \ \hbox{lim}\, \hbox{\boldmath $ |g_B|$} &
  \   \hbox{lim}\, \hbox{\boldmath $|g_{B\!-\!L}|$}\  \ [\,\simeq  \hbox{lim}\, \hbox{$|g_{L}|$}\,]\  &\,  
  \hbox{lim}\, \hbox{\boldmath $|g_{L}|$}\  \ [\,\simeq   \hbox{lim}\, \hbox{$|g_{B\!-\!L}|$}\,]\, 
\\ [2mm]
\, m(\hbox{eV}/c^2) &\, G(x)&
\hbox{(spin-1)}&\hbox{(spin-0)}&\hbox{(spin-1)}\hspace{8mm}\hbox{(spin-0)}\!\!&\ \hbox{(spin-1)}\hspace{6mm}\hbox{(spin-0)}\ \ 
\\ [1mm]
&&&&\\ [-4.5mm]
&&\delta <0 &\delta >0 &\delta >0 &\delta <0 
\\ [2mm]
&&&&\\ [-5.5mm]
&&\propto \, 6\times 1.2  & \propto \,  6&\propto \, 1\ \ \ [\,\hbox{or}\ 1.02\,] &\propto \, 1.02 \times 1.2\ \   [\,\hbox{or}\  1.2\,]
\\ [-4.5mm]
&&&&& \\ [5mm]
&&&&\\ [-8mm]\hline\hline
&&&&\\ [-3mm]
\ \ \ll  10^{-14}\  \ &\simeq 1& \  7.7 \times 10^{-25}  & \  6.4  \times 10^{-25}\,  & \  1.07  \times 10^{-25} \ & \ 1.31  \times 10^{-25} \
\\  [1mm]
&&&& \\ [-5mm]
\ \ 10^{-14} \  \ &1.022 & \ 7.9 \times 10^{-25} & \ 6.6  \times 10^{-25}\,  & \  1.09  \times 10^{-25} \ & \ 1.34  \times 10^{-25} \
\\  [1mm]
&&&&\\ [-5mm]
2\times 10^{-14}\  \ &  1.074& \ 8.3 \times 10^{-25}  & \ 6.9 \times 10^{-25} \, & \  1.15  \times 10^{-25} \ & \ 1.40 \times 10^{-25} \
\\  [1mm]
&&&&\\ [-5mm]
5\times 10^{-14}\ \ &  1.320& \ \ 1.02 \times 10^{-24} \, \ \ & \ 8.5  \times 10^{-25} \, & \  1.41  \times 10^{-25} \ & \ 1.73 \times 10^{-25} \
\\ [5mm]
&&&&\\ [-8mm]\hline
&&&&\\ [-3.5mm]
 \ \ 10^{-13} \ \ &1.886  &1.46\times 10^{-24} \, & \ 1.21\! \times 10^{-24}\  & \ 2.01 \times 10^{-25}\ & 2.47 \times 10^{-25} 
\\ [1mm]
&&&&\\ [-5mm]
2\times10^{-13} \ \ &\, 3.28\ \  &  2.53 \times 10^{-24} \, & \ 2.11\!  \times 10^{-24} \ & \  3.50  \times 10^{-25} \ & \ 4.30 \times 10^{-25} \
\\  [1mm]
&&&&\\ [-5mm]
5\times10^{-13} \ \ &\, 9.43 \ \ & \, 7.3 \times 10^{-24}  & \ 6.1  \times 10^{-24}  & \  1.01 \times 10^{-24} \ & \ 1.23 \times 10^{-24} \
\\ [5mm]
&&&&\\ [-8mm]\hline
&&&&\\ [-3mm]
\ \ 10^{-12}\ \  & 33.8  & 2.61 \times 10^{-23}  & 2.17\! \times 10^{-23} &  3.60 \times 10^{-24} & 4.42 \times 10^{-24} 
\\  [1mm]
&&&&\\ [-5mm]
2\times10^{-12} \ \ & 294\  & \ 2.27 \times 10^{-22} \ & \ 1.89\!  \times 10^{-22} \ & \  3.14  \times 10^{-23} \ & \ 3.84 \times 10^{-23} \
\\  [1mm]
&&&&\\ [-5mm]
5\times10^{-12} \ \ &  1.06\times 10^5 & \ 8.2 \times 10^{-20}  & \ 6.8  \times 10^{-20}  & \  1.13  \times 10^{-20} \ & \ 1.38  \times 10^{-20} \
\\  [1mm]
&&&&\\ [-5mm]
\ \ 10^{-11}\  \ &  1.24\times 10^9& \ 9.6 \times 10^{-16} & \ \ 8\   \times 10^{-16} &  1.33 \times 10^{-16} & 1.63 \times 10^{-16} 
\\ [1.5mm]
\hline
\ea
$
\vspace{4mm}
\end{table}

\vspace{2mm}

\begin{figure}
\caption{\ Upper limits on $|g_{B-L}|$ or  $|g_L| $ (in blue), and $|g_B|$ (in orange), 
depending on the mediator mass $m$ (or range $\lambda = \hbar/mc$), and spin, at the 95\,\% CL.  
The limits for $\delta<0 $ are larger than for  $\delta >0$  by $\simeq \sqrt{6.5/4.5}\,\simeq 1.2\,$. 
For $\lambda \gg R$ they are about $1.1\times 10^{-25}$ for $|g_{B-L}|$ in the spin-1 case or $|g_L|$ in the spin-0 case (solid blue line); 
and $1.3\times 10^{-25}$ for $|g_{B-L}|$ in the spin-0 case or $|g_L|$ in the spin-1 case (dashed blue line).
For $|g_B|$ they are about 6 times larger, at $7.7\times 10^{-25}$ for spin 1  (solid orange) and $6.4 \times 10^{-25}$ for spin-0 (dashed orange). 
\vspace{-.3mm}
\,The limits  increase with $m$ proportionally to 
$ 
e^{mr/2}\,[\,\phi(x)\,(1+mr)\, \bar\rho(x)/\rho_0\,]^{-1/2}$,  behaving at large $x$ like $mR\ e^{mz/2}/\sqrt{1+mr}\,$.
The limits at the 90\,\% CL are slightly lower, down to $10^{-25}$ in the spin-1 case for $B-L$.
\label{lim}}
\vspace{5mm}
\hspace{-10mm}
\ 
\includegraphics[width=8.25cm,height=6.5cm]{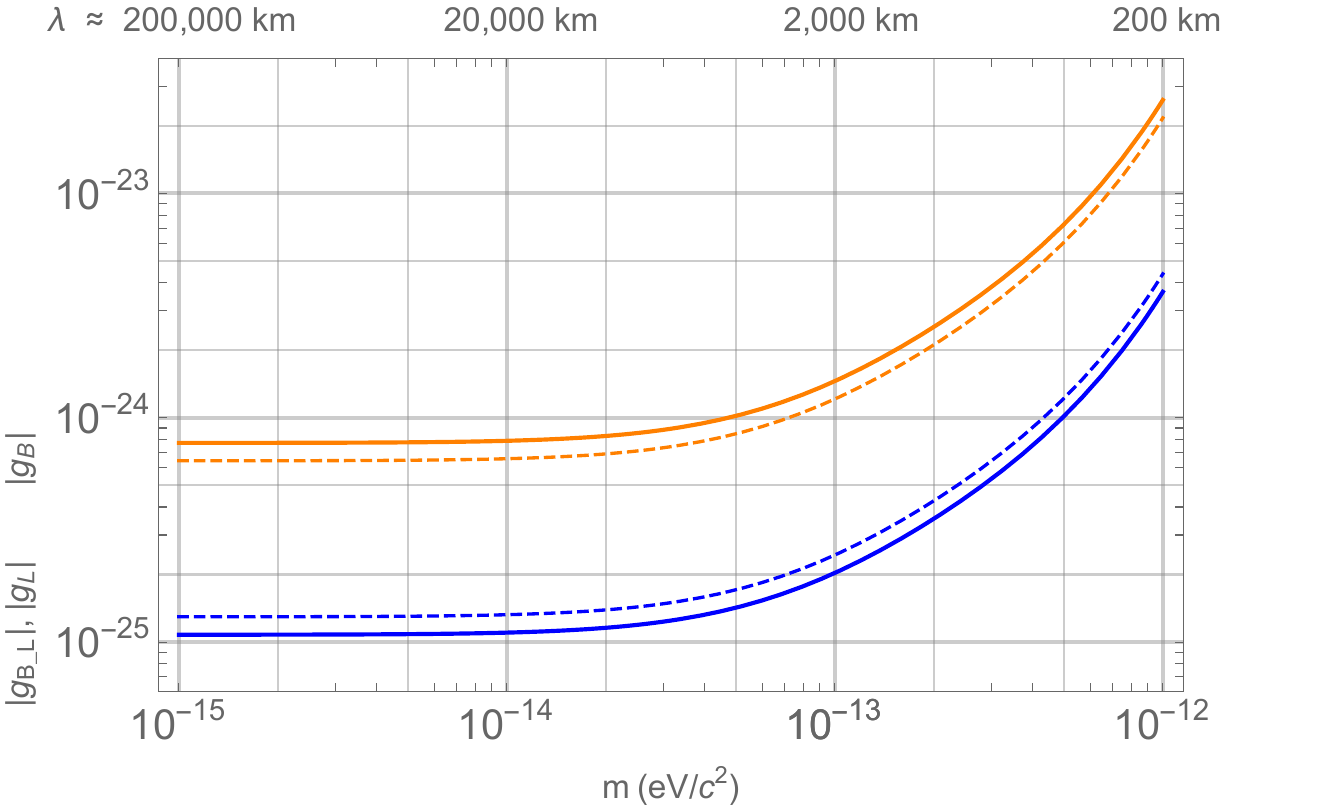} 
\includegraphics[width=8.25cm,height=6.5cm]{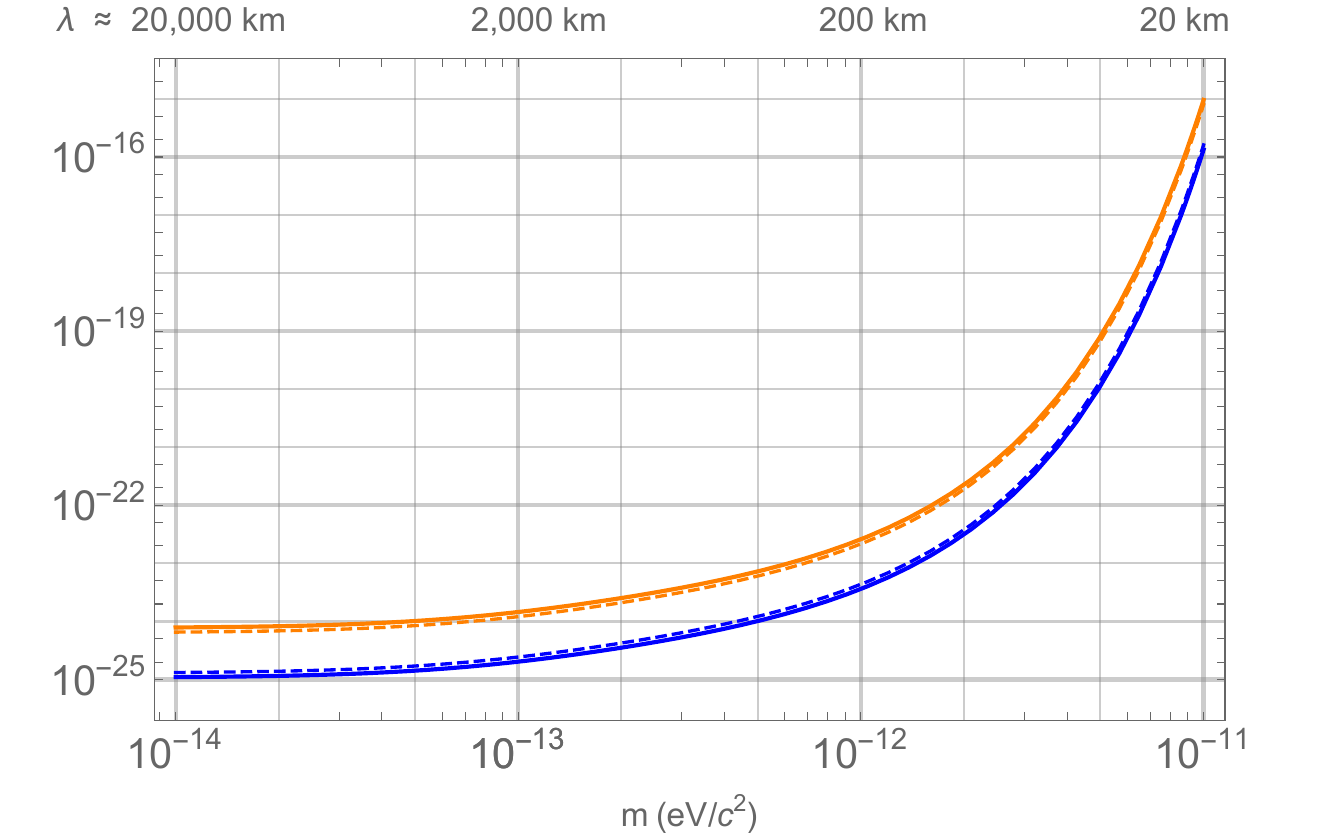}
\hspace{-15mm}
\vspace{1mm}
\end{figure}

The resulting limits on the couplings of a spin-1 or spin-0 particle effectively coupled to $B, \ B-L$, or $L$, are summarized 
in Table \ref{limg} and represented in Fig.\,\ref{lim}, for a mediator mass $m$ from $10^{-15}$ or less, up to $10^{-11}$ eV$/c^2$.

\vspace{1mm}

\section{Conclusions}

The dissymmetry of the final {\it MICROSCOPE}\, result on the E\"otv\"os parameter $\delta$
requires taking into account its sign, leading to
\be
 \delta \,< \,4.5 \times 10^{-15}  \ \  (\delta >0)\ ,\ \ \ \ \ |\delta| < \,6.5  \times 10^{-15}\  \ (\delta <0)\ 
\ \ \ \hbox{(95\,\% CL)}
\ee
(or $3.7 \times 10^{-15}$ and $5.6  \times 10^{-15}$, respectively, at the 90\,\% CL).
For $\delta> 0$ we get almost the same limits (up to $\simeq 2\,\%$)  for a spin-1 coupling to $B\!-\!L$ 
and a spin-0 coupling to $L$; and for $\delta< 0$, for a spin-0 coupling to $B-L$ and a spin 1 coupling to $L$.
\vspace{-.2mm}
The limits for $B$ are about 6 times larger.
For a massless or ultralight mediator we have
\be
\label{bl0}
|g_{B-L}| < \, 
\left\{\, \ba{ccc}1.1\times 10^{-25} & \hbox{(spin 1)}\,, \vspace{2mm}\\ 1.3\times 10^{-25} & \hbox{(spin 0)}\,, \ea  \right.
\ \ |g_B| < \, 
\left\{\, \ba{ccc}7.7\times 10^{-25} & \hbox{(spin 1)}\,, \vspace{2mm}\\ 6.4\times 10^{-25} & \hbox{(spin 0)}\,, \ea  \right.
 \ \ (95\,\%\  \hbox{CL})
\ee
with slightly lower 90\,\% CL limits, including  $10^{-25}$ for a spin-1 coupling to $B\!-\!L$\,.
\vspace{3mm}

The Yukawa potential of a sphere can be expressed in terms of an hyperbolic form factor $\Phi(x\!=\!mR)$. 
We have shown by several methods, including solving the Poisson-like equation for the inside potential,  that it is given by 
\be
\Phi(x) \,=\,\langle\ \frac{\sinh mr}{mr}\ \rangle\ ,
\ee
related by analytic continuation to the form factor
\be
f(k) = \Phi(ikR) \,=\, \langle\ e^{i\,\vec k.\vec r}\ \rangle \,=\,\langle\ \frac{\sin kr}{kr}\ \rangle\ .
\ee
One can also define the hyperbolic form factor
\be
g(\vec k) = f(-i\vec k) = \langle\ e^{\vec k.\vec r}\ \rangle\,=\, \frac{\int \rho(\vec r\,)\  e^{\vec k.\vec r\,}\ d^3\vec r}{\int \rho(\vec r\,) \, d^3\vec r}\,,
\ee
as the bilateral Laplace transform of the distribution $\rho(\vec r\,)$, in general taken with compact support.
For a spherically symmetric distribution $\rho(r)$ normalized to unity,
$\Phi(x) $ may be expressed from the bilateral Laplace transform of $\,r\,\rho(r)$, as
\be
\Phi(x\!=\!kR) = g(k) = \langle\ e^{\vec k.\vec r}\ \rangle 
\,=\,\  \frac{2\pi}{k}\,\int_{-R}^R r\,\rho(r) \ e^{kr}\ dr\,=\,\langle\ \frac{\sinh kr}{kr}\ \rangle\ ,
\ee
reducing to $\phi(x) = 3\, (x \,\cosh x-\sinh x)/x^3$ for an homogeneous sphere. 
An inhomogeneous sphere generates the same outside potential as an homogeneous one with density
$
\bar\rho(x)= \rho_0\,{\Phi(x)}/{\phi(x)}
$, decreasing from the average $\rho_0$ at small $x$ 
down to values representative of an average density around a depth $d \approx \lambda$ (cf. Figs.~\ref{rho} and \,\ref{phiphi}).

\vspace{3mm}

Viewing the Earth as a superposition
of $n$ homogeneous full spheres of radii $R_i$, densities $\Delta \rho_i = \rho_i -\rho_{i+1}$ and masses $m_i$,
one has 
\be
\label{Phicon}
\Phi(x) \,=\, \sum_1^n\ \frac{m_i}{M}\ \phi(mR_i)\,=\,\sum_1^{n-1}\ \frac{\rho_i-\rho_{i+1}}{\rho_0}\ \frac{R_i^3}{R^3} \ \,
\phi\, \hbox{\LARGE $\left(\right.$} x\,\frac{R_i}{R}\hbox{\LARGE $\left.\right)$} \,+\,\frac{\rho_e}{\rho_0}
\ \,\phi(x)\ ,
\ee
\vspace{-3mm}

\noindent
or in the continuum limit
\be
\Phi(x) \,=\, \int_0^R\ \frac{-\,d\rho(r)}{\rho_0}\ \frac{r^3}{R^3} \ \phi\, \hbox{\Large$\left(\right.$} x\,\frac{r}{R}\hbox{\Large$\left.\right)$} \,+\,\frac{\rho_e}{\rho_0}
\ \,\phi(x)\ ,
\ee
evaluated here in a simplified 5-shell model.
$\Phi(x)$ may be developed in terms of the even moments of the density distribution, as
\be
\label{devel}
\ba{ccl}
\Phi(x) &=& \
\,\dis \sum_0^\infty\, \frac{x^{2n}}{(2n+1)!}\, \frac{\langle \,r^{2n}\,\rangle}{R^{2n}} \,=\,
 1\, + \,x^2\,   \frac{\langle\, r^2\,\rangle}{6 \,R^2} +\, x^4\,   \frac{\langle\, r^4\,\rangle}{120\,R^4} +\,x^6\, \frac{\langle \,r^6 \,\rangle}{5\,040\,R^6} 
+ \, ...\ 
\vspace{2mm}\\
&\simeq&\, \dis 1\, + .0827\ x^2 + (2.707 \times 10^{-3})\ x^4 + \,(4.78\times 10^{-5}) \ x^6
+  (5.26\times 10^{-7}) \ x^8
+ ...\ .
 \ea
\ee

\vspace{2mm}

The ratio of a new force induced by a mediator of mass $m$, as compared to a massless one, is
obtained in terms of the effective density $\bar \rho(x) = \rho_0\ \Phi(x)/\phi(x)$, as 
\be
F(x)\, = \,\phi(x)\ (1+mr)\ e^{-mr} \ \,\frac{\bar\rho(x)}{\rho_0}\ ,
\ee
\vspace{-4mm}

\noindent
the coupling limits being  rescaled by $F(x)^{-1/2}$, 
 $\,\simeq 1.9$ or 34 for $m = 10^{-13}$ or $10^{-12}$ eV/$c^2$.
This factor increases very rapidly proportio\-nally to $e^{\,mz/2}$ where $z$ is the satellite altitude, 
up to $\approx 10^9$ at $10^{-11}$ eV/$c^2$ (cf. Table \ref{limg} and Fig.\,\ref{lim}).
In particular, for a spin-1 coupling to $B\!-\!L$, or $B$,
 \be
 \left\{\ \ba{l ccc}
 \hbox{for}\ m \simeq 10^{-13}\  \hbox{eV}/c^2:\, &
|g_{B-L}| < \  \,2 \, \times 10^{-25}\,, &\ |g_B| < 1.5 \,\times 10^{-24}\,, \vspace{2mm}\\
\hbox{for}\ m \simeq 10^{-12}\  \hbox{eV}/c^2:\, & |g_{B-L}| <  \,3.6 \, \times 10^{-24}\,, & \ |g_B| < 2.6\, \times 10^{-23}\, .
\ea \right.
 \ee

\vspace{4mm}

Beyond the specific limits derived here, useful in particular in the context of ultralight dark matter, 
these methods are general, and may be applied to other situations involving finite-range forces, interactions, or phenomena.

\vspace*{2mm}


\end{document}